\documentclass[conference]{IEEEtran}
\usepackage{amsmath}
\usepackage{cite}
\usepackage{graphicx}
\usepackage[table,xcdraw]{xcolor}
\usepackage{bm}
\usepackage{array}
\usepackage{multirow}
\usepackage{makecell}
\usepackage{diagbox}
\usepackage{tabularx}
\usepackage{booktabs}
\usepackage{amssymb}
\usepackage{url}
\usepackage{xcolor}

\begin{document}

\title{A Novel CNN Based Standalone Detector \\for Faster-than-Nyquist Signaling}

\author{
Osman Tokluoglu\IEEEauthorrefmark{1}, Enver Cavus\IEEEauthorrefmark{1}, Ebrahim Bedeer\IEEEauthorrefmark{2}, Halim Yanikomeroglu\IEEEauthorrefmark{3}\\
\IEEEauthorrefmark{1}Department of Electrical and Electronics Engineering, Ankara Yildirim Beyazit University, Ankara, Turkiye\\
\IEEEauthorrefmark{2}Department of Electrical and Computer Engineering, University of Saskatchewan, Saskatoon, SK, Canada\\
\IEEEauthorrefmark{3}Department of System and Computer Engineering, Carleton University, Ottawa, ON, Canada\\
\{otokluoglu, ecavus\}@aybu.edu.tr, e.bedeer@usask.ca, halim@sce.carleton.ca
}

\maketitle

\begin{abstract}
This paper presents a novel convolutional neural network (CNN)-based detector for faster-than-Nyquist (FTN) signaling, introducing structured fixed kernel layers with domain-informed masking to effectively mitigate intersymbol interference (ISI). Unlike standard CNN architectures that rely on moving kernels, the proposed approach employs fixed convolutional kernels at predefined positions to explicitly learn ISI patterns at varying distances from the central symbol. To enhance feature extraction, a hierarchical filter allocation strategy is employed, assigning more filters to earlier layers for stronger ISI components and fewer to later layers for weaker components. This structured design improves feature representation, eliminates redundant computations, and enhances detection accuracy while maintaining computational efficiency. Simulation results demonstrate that the proposed detector achieves near-optimal bit error rate (BER) performance, comparable to the BCJR algorithm for the compression factor $\tau \geq 0.7$, while offering up to $46\%$ and $84\%$ computational cost reduction over M-BCJR for BPSK and QPSK, respectively. Additional evaluations confirm the method’s adaptability to high-order modulations (up to 64-QAM), resilience in quasi-static multipath Rayleigh fading channels, and effectiveness under LDPC-coded FTN transmission, highlighting its robustness and practicality.
\end{abstract}

\begin{IEEEkeywords}
Convolutional Neural Network (CNN), Faster-than-Nyquist (FTN), Intersymbol Interference (ISI) Mitigation, Low-Complexity Detection, Low-Density Parity-Check (LDPC) Coding, Structured Fixed Kernel CNN.
\end{IEEEkeywords}

\section{Introduction}
\label{sec:intro}

The increasing need for high data rates and the scarcity of spectrum resources have necessitated the development of innovative transmission techniques aimed at improving spectral efficiency. One such technique is faster-than-Nyquist (FTN) signaling, which operates beyond the Nyquist rate by intentionally introducing a controlled amount of intersymbol interference (ISI) to boost data transmission rates \cite{mazo1975}. While the Nyquist criterion traditionally guarantees the highest rate without ISI, FTN signaling purposefully violates this principle, leading to ISI and making the detection process more challenging. Although algorithms like BCJR are highly effective for handling the resulting complexity, they demand substantial computational resources \cite{bahl1974}. Consequently, designing low-complexity detection methods that achieve strong performance remains a vital area of research.

In response to this challenge, researchers have explored various techniques aimed at mitigating the ISI introduced by FTN signaling while maintaining computational efficiency. For instance, Prlja and Anderson proposed a reduced-complexity M-BCJR algorithm tailored for turbo equalization in FTN systems \cite{prlja2012}. Their approach efficiently addressed the significant ISI caused by FTN signaling, characterized by the time compression factor $\tau$, which defines the FTN signaling interval relative to the Nyquist interval, across a range of $\tau$ values over additive white Gaussian noise (AWGN) channels. Employing binary and higher-order modulations, they demonstrated that their method achieved bit error rate (BER) performance close to that of the full-complexity BCJR algorithm while reducing computational demands through minimum-phase preprocessing. However, the M-BCJR algorithm still exhibits exponential scaling in complexity with the modulation order, which poses a disadvantage in scenarios requiring higher-order modulations or real-time processing. Similarly, Sugiura et al. introduced a frequency-domain equalization (FDE) technique to mitigate the complexity of time-domain detection in FTN systems \cite{sugiura2013}. Their method, which employs cyclic prefixes and minimum mean square error (MMSE)-based detection, achieved near-optimal BER performance for higher \(\tau\) values when the roll-off factor was \(\beta = 0.5\). However, the reliance on \(\beta = 0.5\) not only limits the understanding of its effectiveness across different roll-off factors but also requires a larger bandwidth, which reduces spectral efficiency and limits its applicability in scenarios where bandwidth is constrained. Additionally, the reliance on cyclic prefixes inherently reduces the spectral efficiency, which is a critical consideration in FTN systems aiming to maximize spectral utilization. A successive symbol-by-symbol sequence estimation (SSSSE) technique for low-complexity FTN signaling detection within the Mazo limit is proposed in \cite{bedeer2017}. Their approach delivered computational efficiency and competitive BER performance, with further enhancements achieved through the implementation of a go-back-\textit{K} strategy to minimize errors. However, the method shows performance limitations at lower \(\tau\) values, such as \(\tau = 0.7\), reducing its effectiveness in scenarios with significant ISI. A simplified M-BCJR algorithm based on the Ungerboeck observation model, incorporating future symbol information to enhance spectral efficiency and signal-to-noise ratio (SNR) for $\tau = 0.8$, is introduced in \cite{li2018}. This method achieves a balance between computational demands and BER performance, demonstrating effective equalization for FTN signaling. A notable limitation of this method is its exponentially increasing computational complexity with higher-order modulation schemes, which, while efficient for binary phase shift keying (BPSK), becomes significantly demanding for modulations like quadrature phase shift keying (QPSK), posing challenges in real-time or resource-constrained applications.

Beyond detection techniques, several studies have investigated transmitter-side precoding strategies to mitigate the ISI introduced by FTN signaling. Wang and Lee \cite{wang1995} proposed an early precoding approach aimed at maximizing the minimum Euclidean distance between FTN sequences. Later, Rusek and Anderson \cite{rusek2008} introduced linear precoding for nonbinary FTN systems to further enhance detection reliability. Pre-equalization methods, including Tomlinson–Harashima precoding and linear spectral precoding, were studied by Jana et al. \cite{jana2017} to simplify receiver complexity. In the context of multiuser systems, Spano et al. \cite{spano2018} developed a spatio-temporal symbol-level precoding framework for FTN in multiuser multiple-input single-output (MISO) channels. Singular value decomposition (SVD)-based precoding with optimal and truncated power allocation was proposed by Ishihara and Sugiura \cite{ishihara2019} to diagonalize the FTN interference. A low-complexity Cholesky decomposition-based linear precoding approach was later introduced by Li et al. \cite{li2021}. Additionally, Liu et al. \cite{liu2023} proposed a joint precoding and pre-equalization scheme tailored for multipath fading channels, while Sugiura \cite{sugiura2021} analyzed the physical-layer security aspects of precoded FTN signaling. Although precoding strategies offer an effective means to mitigate the ISI introduced by FTN signaling, alternative receiver-side detection methods remain a critical research direction.

Motivated by the need for efficient receiver-side detection, and in addition to traditional algorithmic approaches, deep learning has recently emerged as a promising tool for tackling the complexities associated with FTN signaling detection. While many studies integrate deep learning with conventional FTN signaling detection methods to improve interference mitigation, parameter tuning, and overall performance, fewer works employ deep learning architectures as standalone detectors. Baek et al. leveraged long short-term memory (LSTM) networks for FTN signal detection in emergency communication systems, achieving BER performance comparable to the BCJR algorithm for $\tau = 0.8$ while reducing computational complexity \cite{baek2021}. Two deep learning-based receiver architectures: one combining detection with successive interference cancellation (SIC) and another integrating detection and decoding into a single framework are developed in \cite{song2020}. Both approaches achieved near-optimal BER performance for coded and uncoded FTN systems with $\tau = 0.8$, demonstrating resilience to SNR variations. Pan et al. applied a fully connected deep neural network (FC-DNN)-based sliding-window detection method to FTN signaling combined with non-orthogonal multiple access (NOMA) for Internet of Things (IoT) applications. Their approach improved detection accuracy over MMSE-FDE while approaching maximum likelihood performance with reduced latency for $\tau = 0.8$ \cite{pan2020}. A deep learning-enhanced list sphere decoding (DL-LSD) algorithm, leveraging recurrent neural networks (RNNs) and fully connected layers to optimize the initial radius selection process, is proposed in \cite{abbasi2022}. The method reduces computational demands compared to conventional LSD while maintaining strong BER performance for $\tau = 0.8$ and $\tau = 0.9$. Baek et al. further extended deep learning for FTN signaling detection over multipath fading channels by employing bidirectional LSTM (Bi-LSTM) networks, which achieved BER performance comparable to the BCJR algorithm for $\tau = 0.7$ and $\tau = 0.8$, eliminating the need for explicit channel estimation or equalization \cite{baek2023}. Liu et al. proposed a deep learning-assisted sum-product detection algorithm (DL-SPDA) that incorporated a neural network into the conventional sum-product algorithm, significantly reducing residual ISI and addressing short cycle-induced correlations. Their method delivered near-optimal BER performance for $\tau = 0.6$ and $\tau = 0.5$ over AWGN channels with reduced complexity, offering a practical solution for turbo equalization scenarios \cite{liu2021}. In general, these studies either employ deep learning as an auxiliary tool to enhance conventional methods, or focus on RNN-based architectures (particularly LSTM and Bi-LSTM networks), which tend to incur higher computational costs during both training and inference.

While RNNs, such as LSTM networks, have been widely explored for sequence detection in FTN signaling, convolutional neural networks (CNNs) remain largely overlooked despite their potential advantages. Most prior studies have focused on RNN-based architectures due to their ability to capture temporal dependencies, demonstrating strong BER performance. However, these methods often involve high computational complexity and latency, making them less suitable for real-time applications. In contrast, CNNs offer an alternative approach with inherent parallelism and efficient feature extraction, which can significantly reduce computational complexity while maintaining robust detection accuracy. Recent advancements in one-dimensional (1D) CNNs have demonstrated their effectiveness in sequential data processing across various applications \cite{kiranyaz2020review}, such as orthogonal frequency division multiplexing (OFDM) systems, where 1D CNNs have replaced traditional receiver modules and improved performance under diverse channel conditions \cite{WANG2023102055}. Hybrid architectures that combine 1D CNNs with recurrent models, such as Bi-LSTM networks, have further illustrated the potential of CNNs in addressing ISI challenges in FTN signaling \cite{yang2024mhsa}. A recent study by De Filippo et al. explores the application of CNNs for FTN detection, employing skip connections within a conventional CNN framework to enhance feature extraction \cite{defilippo2025}. Their work demonstrates that deep learning can effectively mitigate ISI, achieving notable performance improvements over traditional equalization techniques. This study further highlights the potential of CNNs in FTN detection, reinforcing the advantages of deep feature extraction in handling challenging communication scenarios. While their approach successfully leverages CNN architectures, it primarily relies on standard convolutional layers without explicitly incorporating domain knowledge of ISI characteristics. Instead, feature extraction is performed through deep stacked convolutional layers, which may introduce computational overhead and limit interpretability in ISI-dominated environments. These studies highlight the capability of 1D CNNs to efficiently learn temporal dependencies without the need for complex preprocessing, making them a compelling alternative to traditional approaches.

Recognizing this potential, this work proposes a standalone CNN-based FTN detector that leverages structured fixed-kernel layers with domain-informed masking to provide an efficient and effective solution for FTN signal detection under varying ISI conditions. The study focuses on FTN signaling with $\tau \geq 0.7$, a regime that offers a favorable trade-off between spectral efficiency and computational complexity, where the proposed low-complexity method demonstrates its highest effectiveness. Unlike standard CNN architectures, which rely on moving kernels, the proposed approach employs fixed convolutional kernels at predefined positions to explicitly learn ISI patterns at varying distances. To further improve feature extraction, a hierarchical filter allocation strategy is introduced, assigning more filters to the initial kernel layers to capture stronger ISI effects, while subsequent layers use fewer filters to process weaker ISI components. This structured methodology enhances feature representation, reduces redundant computations, and improves detection accuracy while maintaining computational efficiency. Simulation results demonstrate that the proposed detector achieves near-optimal BER performance, comparable to the BCJR algorithm for compression factors $\tau \geq 0.7$, while achieving up to $46\%$ computational efficiency improvement over the M-BCJR algorithm for BPSK modulation and up to $84\%$ improvement for QPSK modulation. The BER performance and computational complexity of the proposed approach are also analyzed in comparison with other existing methods in the literature, underscoring its efficiency and practicality. Furthermore, the proposed detector demonstrates strong resilience under quasi-static multipath fading channels, achieving near-optimal BER performance comparable to the BCJR algorithm even in the presence of severe channel impairments. In addition, the method is tested with higher-order modulations up to 64-QAM, showcasing its capability to handle such modulations. Moreover, to validate its applicability to practical coded systems, the proposed detector is also evaluated under Low-Density Parity-Check (LDPC)-coded FTN transmission scenarios, demonstrating reliable performance improvements in conjunction with forward error correction. These findings establish the CNN-based fixed-kernel approach as an efficient and accurate alternative for FTN signaling detection, offering a balance between performance and complexity across diverse operating conditions.

The remainder of this paper is organized as follows: Section~\ref{sec:system_model} describes the system model for FTN signaling, providing an overview of the signal structure and interference modeling. Section~\ref{sec:cnn_tech} details the proposed CNN-based FTN detector, including its architecture, data preprocessing techniques, and training strategies. In Section~\ref{sec:Simulation Results}, the simulation results are presented, with a focus on BER performance and computational efficiency comparisons. Finally, Section~\ref{sec:conclusion} concludes the paper with a summary of the findings and potential directions for future work.

\section{System Model}
\label{sec:system_model}

For a baseband pulse \(g(t)\) with a bandwidth of \((1 + \beta)/2T\), where \( \beta \) is the roll-off factor, the transmitted signal can be expressed as
\begin{equation}
s(t) = \sum_{k} a_k g(t - k{\tau}T).
\label{eq:transmission_signal}
\end{equation}

In this expression  \( s(t) \) represents the transmitted signal, $a_k$ represents the $k$-th data symbol, which is modulated using either BPSK or higher order modulations up to 64-QAM. The parameter $T$ is the Nyquist symbol duration, corresponding to the time interval between consecutive symbols in the absence of compression. The compression factor \( \tau \), where \( 0 < \tau < 1 \), determines the signaling rate relative to the Nyquist rate. A smaller value of \( \tau \) leads to a higher transmission rate by accelerating the symbol placement by a factor of \( 1/\tau \), which enhances spectral efficiency. However, reducing \( \tau \) below 1 results in overlapping pulses that introduce ISI, as the bandwidth of the pulse \( g(t) \) is \((1 + \beta)/2T\), which is narrower than \((1 + \beta)/2\tau T\). This overlap causes symbols to interfere with one another, complicating signal detection at the receiver. Effective detection techniques must compensate for this ISI to ensure accurate symbol recovery. Additionally, the pulse \( g(t) \) is normalized to ensure unit energy, defined as

\begin{equation}
\int_{-\infty}^{+\infty} |g(t)|^2 \, dt = 1.
\end{equation}

In practical applications, a root-raised cosine (RRC) pulse with an excess bandwidth of $\beta = 0.35$ is frequently employed, as it is a standard choice in the field \cite{benedetto1993}. The spectral efficiency improvement is quantified by $1 / \tau$. For example, when $\tau = 0.9$, the corresponding gain in spectral efficiency is approximately $11\%$.

Firstly, this scenario is analyzed within the framework of an AWGN channel, characterized by a noise distribution of $\cal{N}$ $(0, N_0 / 2)$. After passing through the AWGN channel and undergoing post-filtering, the received signal can be represented as
\begin{equation}
y(n\tau T) = \sum_{k} a_kx((n - k)\tau T) + w(n\tau T), 
\label{eq:received_signal}
\end{equation}
where the signal $x(t)$ is defined as the convolution of the transmit filter $g(t)$ with its time-reversed version $g(-t)$, expressed as
\begin{equation}
x(t) = g(t) * g(-t),
\end{equation}
where $*$ denotes the convolution operation and $w(n\tau T)$ represents the $n$th colored noise sample.

\begin{figure}[!t]
\centering
\includegraphics[width=\linewidth]{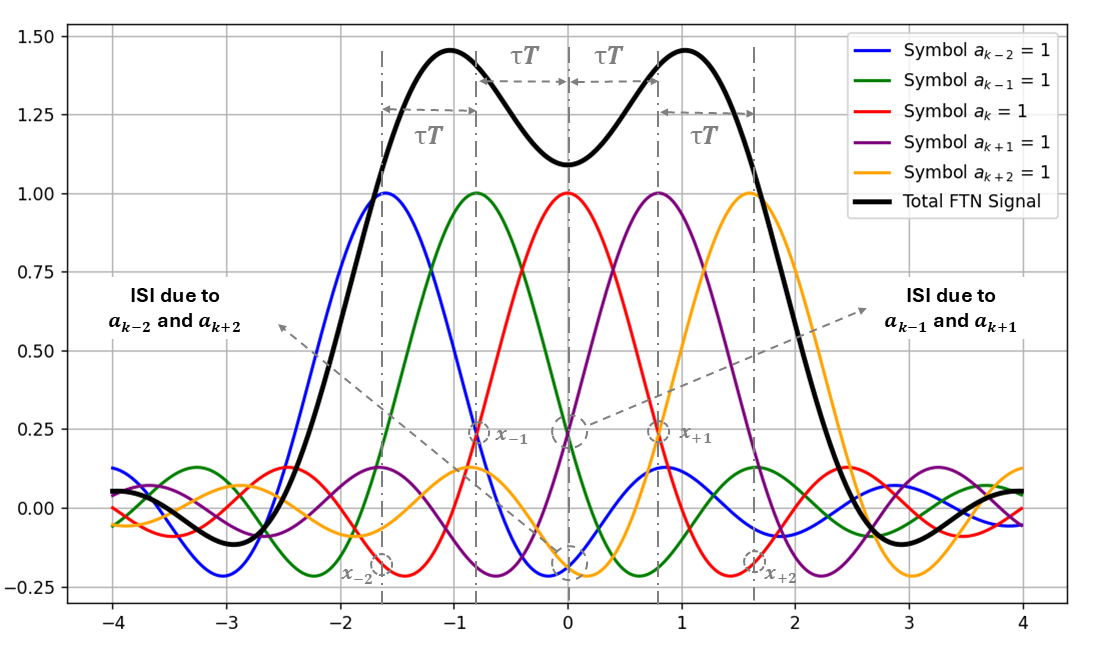}
\caption{Faster-than-Nyquist signaling structure and interference coefficients.}
\label{fig:Fig1_FTN_Structure}
\end{figure}

The structure of the FTN signal and the corresponding interference coefficients are depicted for a sequence of five BPSK symbols
$ 
\begin{bmatrix}
1 & 1 & 1 & 1 & 1
\end{bmatrix}
$
in Fig.~\ref{fig:Fig1_FTN_Structure}. When detecting the $k$-th symbol, represented as $a_k$, interference arises from the symbols immediately preceding and succeeding it, specifically $a_{k-2}$, $a_{k-1}$, $a_{k+1}$, and $a_{k+2}$. The number of interfering symbols, denoted by $N$, corresponds to the one-sided ISI length and is determined by the value of $\tau$. In Fig.~\ref{fig:Fig1_FTN_Structure}, $N$ is illustrated as 2, the black curve represents the total transmitted signal, which includes ISI contributions from neighboring symbols, while the colored curves correspond to the individual contributions of each transmitted symbol. The filter coefficients are additionally shown in Fig.~\ref{fig:Fig1_FTN_Structure} as
$ 
\begin{bmatrix}
x_{-2} & x_{-1} & x_{+1} & x_{+2}
\end{bmatrix}
$,
where each $x$ value corresponds to the coefficient associated with the interfering symbols for $\tau = 0.8$. The coefficients $x_{-1}$ and $x_{+1}$ represent the first-order interference coefficients and are identical. Similarly, $x_{-2}$ and $x_{+2}$ correspond to the second-order interference coefficients and are also equal. Using this representation, the received signal $y_k$ for the $k$-th symbol can be expressed mathematically as

\begin{equation}
\scalebox{0.85}{%
$\displaystyle
y_k = 
\big[ x_N \, x_{N-1} \, \dots \, x_1 \, x_0 \, x_1 \, \dots \, x_{N-1} \, x_N \big]
\begin{bmatrix}
a_{k-N} \\ 
a_{k-N+1} \\ 
\vdots \\ 
a_{k-1} \\ 
a_k \\ 
a_{k+1} \\ 
\vdots \\ 
a_{k+N-1} \\ 
a_{k+N}
\end{bmatrix}
+ w_k.$%
}
\label{eq:received_signal}
\end{equation}

As an illustrative example, a transmitted sequence with a length of $K = 7$ and $N = 2$ can be analyzed. In this scenario, the complete received signal is represented in vector form as

\begin{multline}
\scalebox{0.9}{$\displaystyle
\begin{bmatrix}
y_0 \\ 
y_1 \\ 
y_2 \\ 
y_3 \\ 
y_4 \\ 
y_5 \\ 
y_6
\end{bmatrix}
=
\begin{bmatrix}
x_0 & x_1 & x_2 & 0 & 0 & 0 & 0 \\ 
x_1 & x_0 & x_1 & x_2 & 0 & 0 & 0 \\ 
x_2 & x_1 & x_0 & x_1 & x_2 & 0 & 0 \\ 
0 & x_2 & x_1 & x_0 & x_1 & x_2 & 0 \\ 
0 & 0 & x_2 & x_1 & x_0 & x_1 & x_2 \\ 
0 & 0 & 0 & x_2 & x_1 & x_0 & x_1 \\ 
0 & 0 & 0 & 0 & x_2 & x_1 & x_0
\end{bmatrix}
\begin{bmatrix}
a_0 \\ 
a_1 \\ 
a_2 \\ 
a_3 \\ 
a_4 \\ 
a_5 \\ 
a_6
\end{bmatrix}
+
\begin{bmatrix}
w_0 \\ 
w_1 \\ 
w_2 \\ 
w_3 \\ 
w_4 \\ 
w_5 \\ 
w_6
\end{bmatrix}
.$}
\label{eq:matrix_full}
\end{multline}

In this context, the received signal $\mathbf{y}$ is obtained by multiplying the data symbols with the corresponding filter coefficients. Consequently, the general expression for this relationship is given in (\ref{eq:matrix_form}) as
\begin{multline}
\mathbf{y} = \mathbf{X} \mathbf{a} + \mathbf{w}, \quad
\\
\scriptsize
\hspace*{-29.3em}
\begin{bmatrix}
y_0 \\ 
\vdots \\ 
y_{K-1}
\end{bmatrix}
=
\\ 
\scriptsize
\begin{bmatrix}
x_0 & x_1 & x_2 & \cdots & x_N & 0 & \cdots & 0 \\
x_1 & x_0 & x_1 & \cdots & x_{N-1} & x_N & \cdots & 0 \\
x_2 & x_1 & x_0 & \cdots & x_{N-2} & x_{N-1} & \cdots & 0 \\
\vdots & \vdots & \vdots & \ddots & \vdots & \vdots & \ddots & \vdots \\
x_N & x_{N-1} & x_{N-2} & \cdots & x_0 & x_1 & \cdots & x_N \\
0 & x_N & x_{N-1} & \cdots & x_1 & x_0 & \cdots & x_{N-1} \\
0 & 0 & x_N & \cdots & x_2 & x_1 & \cdots & x_{N-2} \\
\vdots & \vdots & \vdots & \ddots & \vdots & \vdots & \ddots & \vdots \\
0 & 0 & 0 & \cdots & x_N & x_{N-1} & \cdots & x_0
\end{bmatrix}
\begin{bmatrix}
a_0 \\
a_1 \\
a_2 \\
a_3 \\
\vdots \\
\\
a_{K-4} \\
a_{K-3} \\
a_{K-2} \\
a_{K-1}
\end{bmatrix}
\\
+
\scriptsize
\begin{bmatrix}
w_0 \\ 
\vdots \\ 
w_{K-1}
\end{bmatrix},
\label{eq:matrix_form}
\end{multline}
where $\mathbf{y}$ represents the received samples vector, $\mathbf{a}$ is the $K$x$1$ transmitted data symbols vector, $\mathbf{w}$ $\mathtt{\sim}$ $\cal{N}$ $(0, \mathbf{X}N_0 / 2)$ is the $K$x$1$ Gaussian noise samples, with  $\mathbf{X}$ being the $K$x$K$ ISI matrix. The precise values of the ISI coefficients, ranging from $x_0$ to $x_8$, for $\tau$ values of 0.7, 0.8, and 0.9 are presented in Table~\ref{tab:isi_coefficients}. As shown in the table, a decrease in the parameter $\tau$ leads to more pronounced ISI effects on the estimated symbol, reflecting greater symbol overlap in the time domain. For a more detailed exposition of FTN signaling theory, including the derivation of ISI coefficients and the impact of the compression factor~$\tau$, readers are referred to foundational and survey works such as~\cite{mazo1975, anderson2013, dasalukunte2014, ishihara2021}.

\begin{table}[h!]
\caption{One-sided ISI coefficients for $\tau = 0.7$, $0.8$, and $0.9$.}
\centering
\renewcommand{\arraystretch}{1.5} % Adjust row height
\setlength{\tabcolsep}{6pt} % Adjust column spacing
\begin{tabular}{>{\centering\arraybackslash}m{1.3cm} c c c}
\rowcolor[gray]{0.9}
\textbf{Coefficient} & \textbf{$\tau = 0.7$} & \textbf{$\tau = 0.8$} & \textbf{$\tau = 0.9$} \\ 
$x_0$ & $9.99 \times 10^{-1}$ & $9.99 \times 10^{-1}$ & $9.99 \times 10^{-1}$ \\ 
\rowcolor[gray]{0.95}
$x_1$ & $3.53 \times 10^{-1}$ & $2.22 \times 10^{-1}$ & $1.02 \times 10^{-1}$ \\ 
$x_2$ & $-1.83 \times 10^{-1}$ & $-1.52 \times 10^{-1}$ & $-7.86 \times 10^{-2}$ \\ 
\rowcolor[gray]{0.95}
$x_3$ & $3.24 \times 10^{-2}$ & $7.62 \times 10^{-2}$ & $4.87 \times 10^{-2}$ \\ 
$x_4$ & $3.16 \times 10^{-2}$ & $-2.44 \times 10^{-2}$ & $-2.26 \times 10^{-2}$ \\ 
\rowcolor[gray]{0.95}
$x_5$ & $-2.79 \times 10^{-2}$ & $5.93 \times 10^{-3}$ & $1.24 \times 10^{-2}$ \\ 
$x_6$ & $1.37 \times 10^{-2}$ & $3.16 \times 10^{-3}$ & $-3.61 \times 10^{-3}$ \\ 
\rowcolor[gray]{0.95}
$x_7$ & $3.31 \times 10^{-4}$ & $-1.73 \times 10^{-3}$ & $9.81 \times 10^{-4}$ \\ 
$x_8$ & $-1.73 \times 10^{-3}$ & $6.64 \times 10^{-4}$ & $-1.69 \times 10^{-4}$ \\ 
\end{tabular}
\label{tab:isi_coefficients}
\end{table}

In practical wireless environments, the transmitted signal is not only affected by additive noise but also by multipath fading, where the signal reaches the receiver through multiple propagation paths. Each path introduces a delay and attenuation due to effects such as reflection, scattering, and diffraction. To model this, the received signal is extended to include multiple channel taps as follows:

\begin{equation}
    y(n\tau T) = \sum_{l=0}^{L-1} \sum_{k} h_l a_k x((n - k - \delta_l)\tau T) + w(n\tau T).
\end{equation}

Here, $a_k$ denotes the $k$-th transmitted symbol, and $h_l$ represents the channel coefficient associated with the $l$-th multipath component. The total number of multipath components is denoted by $L$. Each path introduces a relative delay $\delta_l$ (in symbols) and contributes a delayed and scaled version of the signal. The overall received signal is then formed by the superposition of these components, each arriving with its own delay and amplitude. The complete channel impulse response can thus be expressed as
\begin{equation}
    \mathbf{h} = [h_0, h_1, \ldots, h_{L-1}].
\end{equation}

\section{CNN Based FTN Signaling Detection Technique}
\label{sec:cnn_tech}

\begin{figure}[!t]
\centering
\includegraphics[width=\linewidth]{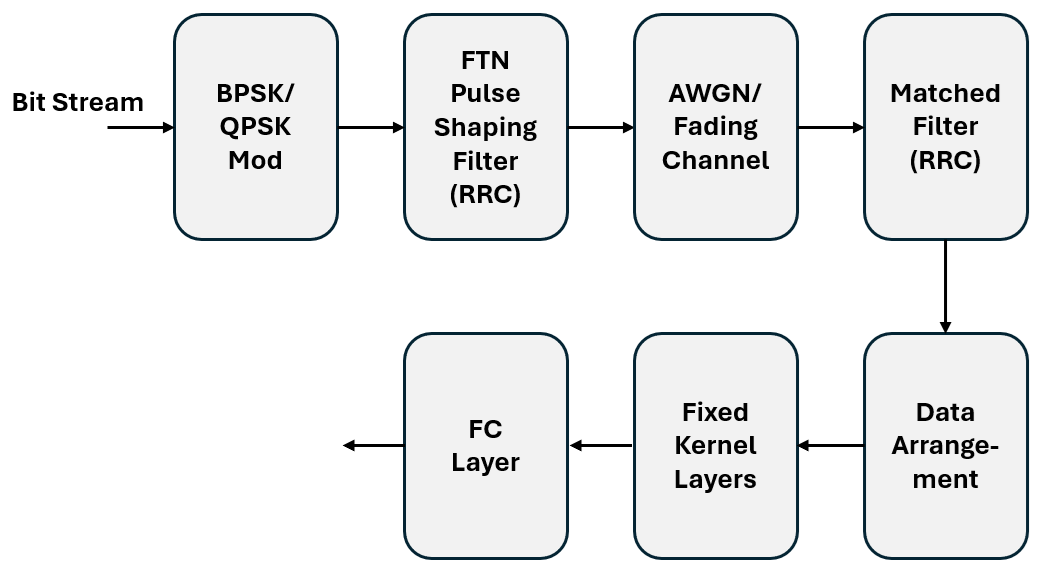}
\caption{FTN signaling-based communication system with proposed signal detector.}
\label{fig:Fig2_Com_Sys_Arch}
\end{figure}

In this section, we present the methodology underlying the proposed CNN-based FTN detector. The development process encompasses multiple components, including dataset preparation, data arrangement, system modeling, and optimization strategies, each carefully designed to address the challenges associated with ISI in FTN signaling. By leveraging structured fixed kernel layers and domain-specific preprocessing techniques, the proposed methodology systematically captures the ISI effects introduced by FTN signaling, enabling accurate symbol detection across varying $\tau$ values. FTN signaling-based communication system with the proposed signal detector is depicted in Fig.~\ref{fig:Fig2_Com_Sys_Arch}. Additionally, the computational efficiency of the structured approach is analyzed and compared with conventional methods, highlighting the advantages of the proposed design. The details of each component are outlined in the following subsections.

\subsection{Dataset Preparation}
\label{subsec:dataset}

To facilitate the training of the proposed deep learning-based FTN detector, a comprehensive dataset was generated, covering a wide range of $\tau$ and SNR values within a BPSK- and higher-order modulated FTN communication system, including up to 64-QAM, operating over an AWGN channel. The dataset comprises input-output pairs for each unique $\tau$ and SNR combination, ensuring diverse training scenarios to improve model generalization across varying ISI conditions and noise levels.

To enhance the neural network’s ability to learn the ISI characteristics inherent to FTN signaling, datasets with high SNR values were emphasized during training for each  value. This approach enables the model to effectively learn and capture the underlying ISI patterns and dynamics under high SNR conditions, facilitating a deeper understanding of ISI effects before inference. Extensive training at very low SNRs was intentionally avoided, as the resulting BER performance in such regimes is typically poor across all detectors and does not reflect the intended operating region of FTN-based systems. However, additional testing under low SNR conditions was conducted to ensure robustness and assess the model's adaptability to practical deployment scenarios. Furthermore, a separate model was trained for each $\tau$ value, as ISI patterns vary significantly with the compression factor. This approach enables the CNN-based FTN detector to learn $\tau$-specific signal characteristics more accurately, improving detection performance and avoiding generalization errors observed when using a single model across different $\tau$ values.

\subsection{Data Arrangement}
\label{subsec:data_arrangement}

In FTN signaling, the ISI is influenced by both preceding and succeeding symbols relative to the symbol of interest. The extent of this interference is determined by the value of $\tau$. To account for this, the system architecture incorporates the range of symbols contributing to interference. During both training and inference, the value of a specific symbol is estimated using $N$ symbols from its preceding and succeeding neighbors, corresponding to the one-sided ISI length. To streamline this process, the input data is preprocessed to group each target symbol with its $N$ neighboring symbols on both sides, rather than feeding the symbols individually. Specifically, $2N+1$ received symbols are used to predict a single transmitted symbol, as
\begin{equation}
\mathbf{y}_k = 
\begin{bmatrix}
y_{k-N} & y_{k-N+1} & \cdots & y_k & \cdots & y_{k+N-1} & y_{k+N}
\end{bmatrix},
\label{eq:yk_vector}
\end{equation}
where $y_k$ represents the $k$-th received symbol corresponding to the $k$-th transmitted symbol $a_k$. The Data Arrangement block organizes the received symbols as shown in (\ref{eq:yk_vector}). However, for the first $N$ and last $N$ symbols of the dataset, insufficient neighboring symbols are available to fill the required range. To address this, the missing positions are padded with zeroes.

A CNN-based FTN detector was implemented to enhance detection accuracy across different $\tau$ values. The parameter $N$ was selected based on empirical observations derived from extensive simulations and the one-sided ISI coefficients provided in Table~\ref{tab:isi_coefficients}. For instance, $N=2$ was chosen for $\tau=0.9$, $N=6$ for $\tau=0.8$, and $N=8$ for $\tau=0.7$, achieving an optimal trade-off between computational efficiency and detection performance. Increasing $N$ improves the model’s ability to capture ISI effects but also increases computational costs, which were carefully considered during selection. For $N=6$, the data structure layout is shown in Table~\ref{table:ftn_data_layout_N=6}, where each row is provided as input to the CNN-based detection system.

\begin{table}[ht]
\centering
\caption{Data Structure for $N=6$}
\label{table:ftn_data_layout_N=6}
\renewcommand{\arraystretch}{1.3} % Adjust row spacing
\setlength{\tabcolsep}{2pt} % Reduce column spacing
\resizebox{\columnwidth}{!}{ % Resize table to fit within column width
\begin{tabular}{cccccccccccccccc}
0th input   & 0 & 0 & 0 & 0 & 0 & 0 & \bm{$y_0$} & $y_1$ & $y_2$ & $y_3$ & $y_4$ & $y_5$ & $y_6$ \\ 
\rowcolor[HTML]{EFEFEF} 
1st input   & 0 & 0 & 0 & 0 & 0 & $y_0$ & \bm{$y_1$} & $y_2$ & $y_3$ & $y_4$ & $y_5$ & $y_6$ & $y_7$ \\ 
2nd input   & 0 & 0 & 0 & 0 & $y_0$ & $y_1$ & \bm{$y_2$} & $y_3$ & $y_4$ & $y_5$ & $y_6$ & $y_7$ & $y_8$ \\ 
\rowcolor[HTML]{EFEFEF} 
3rd input   & 0 & 0 & 0 & $y_0$ & $y_1$ & $y_2$ & \bm{$y_3$} & $y_4$ & $y_5$ & $y_6$ & $y_7$ & $y_8$ & $y_9$ \\ 
4th input   & 0 & 0 & $y_0$ & $y_1$ & $y_2$ & $y_3$ & \bm{$y_4$} & $y_5$ & $y_6$ & $y_7$ & $y_8$ & $y_9$ & $y_{10}$ \\ 
\rowcolor[HTML]{EFEFEF} 
5th input   & 0 & $y_0$ & $y_1$ & $y_2$ & $y_3$ & $y_4$ & \bm{$y_5$} & $y_6$ & $y_7$ & $y_8$ & $y_9$ & $y_{10}$ & $y_{11}$ \\ 
6th input   & $y_0$ & $y_1$ & $y_2$ & $y_3$ & $y_4$ & $y_5$ & \bm{$y_6$} & $y_7$ & $y_8$ & $y_9$ & $y_{10}$ & $y_{11}$ & $y_{12}$ \\ 
\rowcolor[HTML]{EFEFEF} 
$\vdots$    & $\vdots$ & $\vdots$ & $\vdots$ & $\vdots$ & $\vdots$ & $\vdots$ & $\vdots$ & $\vdots$ & $\vdots$ & $\vdots$ & $\vdots$ & $\vdots$ & $\vdots$ \\ 
$(k-6)$th input   & $y_{k-12}$ & $y_{k-11}$ & $y_{k-10}$ & $y_{k-9}$ & $y_{k-8}$ & $y_{k-7}$ & \bm{$y_{k-6}$} & $y_{k-5}$ & $y_{k-4}$ & $y_{k-3}$ & $y_{k-2}$ & $y_{k-1}$ & $y_{k}$ \\
\rowcolor[HTML]{EFEFEF} 
$(k-5)$th input   & $y_{k-11}$ & $y_{k-10}$ & $y_{k-9}$ & $y_{k-8}$ & $y_{k-7}$ & $y_{k-6}$ & \bm{$y_{k-5}$} & $y_{k-4}$ & $y_{k-3}$ & $y_{k-2}$ & $y_{k-1}$ & $y_{k}$ & 0 \\
$(k-4)$th input   & $y_{k-10}$ & $y_{k-9}$ & $y_{k-8}$ & $y_{k-7}$ & $y_{k-6}$ & $y_{k-5}$ & \bm{$y_{k-4}$} & $y_{k-3}$ & $y_{k-2}$ & $y_{k-1}$ & $y_{k}$ & 0 & 0 \\
\rowcolor[HTML]{EFEFEF} 
$(k-3)$th input   & $y_{k-9}$ & $y_{k-8}$ & $y_{k-7}$ & $y_{k-6}$ & $y_{k-5}$ & $y_{k-4}$ & \bm{$y_{k-3}$} & $y_{k-2}$ & $y_{k-1}$ & $y_{k}$ & 0 & 0 & 0 \\
$(k-2)$th input   & $y_{k-8}$ & $y_{k-7}$ & $y_{k-6}$ & $y_{k-5}$ & $y_{k-4}$ & $y_{k-3}$ & \bm{$y_{k-2}$} & $y_{k-1}$ & $y_{k}$ & 0 & 0 & 0 & 0 \\
\rowcolor[HTML]{EFEFEF} 
$(k-1)$th input   & $y_{k-7}$ & $y_{k-6}$ & $y_{k-5}$ & $y_{k-4}$ & $y_{k-3}$ & $y_{k-2}$ & \bm{$y_{k-1}$} & $y_{k}$ & 0 & 0 & 0 & 0 & 0 \\
$k$th input   & $y_{k-6}$ & $y_{k-5}$ & $y_{k-4}$ & $y_{k-3}$ & $y_{k-2}$ & $y_{k-1}$ & \bm{$y_{k}$} & 0 & 0 & 0 & 0 & 0 & 0 \\ 
\end{tabular}
}
\end{table}

\subsection{CNN-Based FTN Detector System Model}
\label{subsec:cnn_model}

The key challenge in designing an effective CNN-based FTN detector lies in the accurate modeling of ISI, which is induced by the compressed symbol intervals defined by the parameter \( \tau \). As detailed in Section~\ref{sec:system_model}, the received symbol at discrete time \(k\) can be modeled as

\begin{equation}
y_k = \sum_{i=-N}^{N} x_i a_{k-i} + w_k,
\label{eq:cnn_isi_model}
\end{equation}
where \( a_{k-i} \in \mathcal{A} \) denotes the transmitted symbol drawn from a modulation alphabet \(\mathcal{A}\), \( x_i \) is the ISI coefficient at lag \( i \), \( w_k \sim \mathcal{N}(0, \sigma^2) \) represents colored Gaussian noise, and \( N \) is the one-sided ISI length determined by \( \tau \). This formulation highlights that each observation \( y_k \) is influenced by a linear combination of \( 2N + 1 \) transmitted symbols. While this model is initially formulated under the AWGN channel, the proposed CNN-based architecture is later evaluated under both AWGN and multipath fading environments, as discussed in Sections~\ref{subsec:ber_results} and~\ref{subsection:last_comp}.

To exploit this structure, the input to the CNN detector is constructed as a vector \(\mathbf{y}_k \in \mathbb{R}^{2N+1}\), defined as in~(\ref{eq:yk_vector}).

Our initial investigation focused on applying conventional convolutional kernels to the input vector \( \mathbf{y}_k \in \mathbb{R}^{2N+1} \), which comprises the received symbol \( y_k \) and its \( N \) adjacent neighbors on either side. In this approach, a learnable convolutional kernel \( \boldsymbol{\theta} \in \mathbb{R}^{r} \), where \( r \leq 2N + 1 \), is applied to extract local features from the input. The convolution operation at position \( m \), corresponding to the \( m \)-th valid sliding window over the input sequence, is computed as
\begin{equation}
z_m = \sum_{j=0}^{r-1} \theta_j y_{k - N + m + j}, \quad m \in [0, 2N - r + 1],
\label{eq:conv_standard}
\end{equation}
where \( z_m \) denotes the pre-activation output at position \( m \), \( \theta_j \) is the \( j \)-th element of the convolutional kernel, and \( y_{k - N + m + j} \) is the corresponding input sample from the local context window centered at symbol index \( k \). Here, \( r \) is the kernel size and \( m \) ranges over all positions where the kernel fully fits within the input window of size \( 2N + 1 \). The activated output is obtained as

\begin{equation}
o_m = \phi(z_m) = \phi\left( \sum_{j=0}^{r-1} \theta_j y_{k - N + m + j} \right),
\end{equation}
where \( \phi(\cdot) \) denotes a nonlinear activation function, such as \texttt{ReLU} or \texttt{tanh}.

\begin{figure}[ht]
    \centering
    \includegraphics[width=\columnwidth]{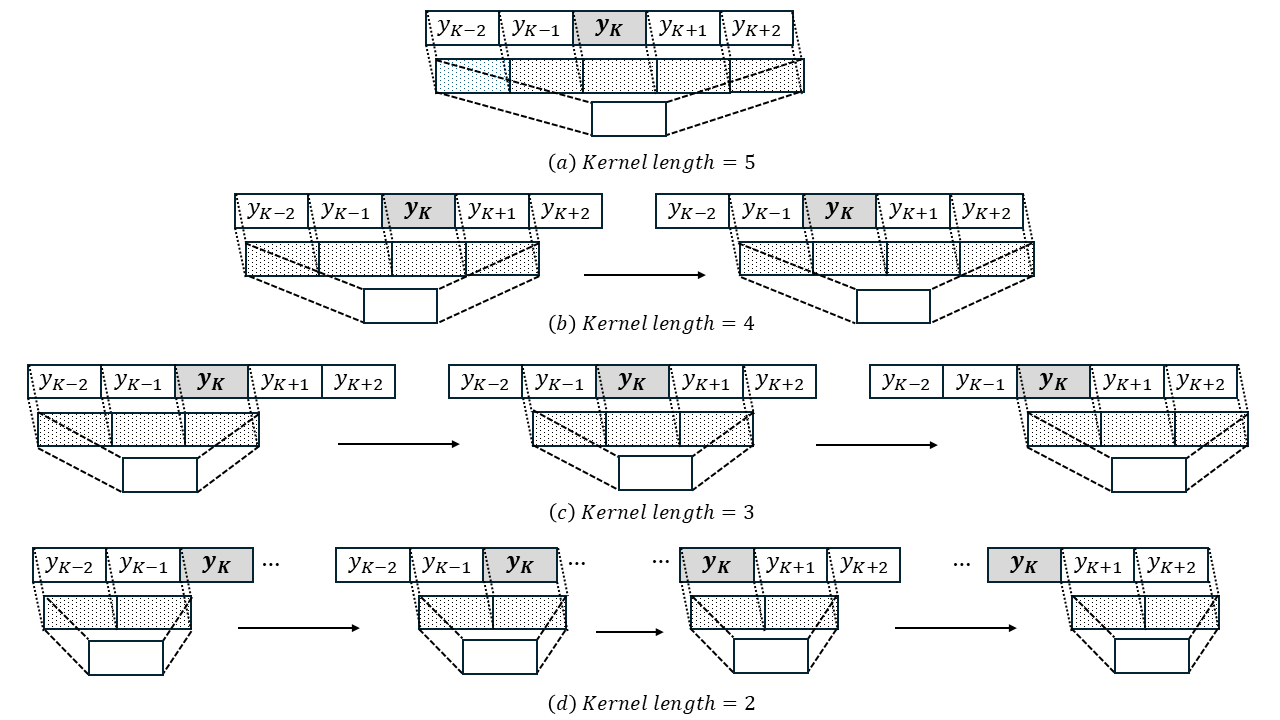}
    \caption{Illustration of conventional kernelization techniques with varying kernel sizes (\(K = 5\), \(N = 2\)): (a) 5, (b) 4, (c) 3, and (d) 2.}
    \label{fig:Fig3_Conventional_Kernels}
\end{figure}

While this technique captures local correlations, it has two major limitations in the FTN context. First, larger kernel sizes (\( r \approx 2N + 1 \)) aggregate broader context at the cost of diluting strong ISI-induced dependencies near \( y_k \). Second, smaller kernels (\( r \ll 2N + 1 \)) may fail to span the ISI region or align with the central symbol, degrading detection accuracy. These limitations, illustrated in Fig.~\ref{fig:Fig3_Conventional_Kernels}, motivated the development of a more structured approach based on fixed kernel layers and domain-informed masking.

To address the limitations of conventional kernels, we propose \textit{structured fixed kernel layers} where each kernel focuses on a predefined positional relationship relative to the central symbol. Specifically, for a given ISI span \(N\), we define \(N\) fixed kernel layers, each designed to focus on received symbols at a distance \(i \in \{1, 2, \dots, N\}\) from the central position \(y_k\). Each kernel processes a triplet consisting of the central symbol and its symmetric neighbors as

\begin{equation}
\mathbf{y}_k^{(i)} = 
\begin{bmatrix}
y_{k-i} & y_k & y_{k+i}
\end{bmatrix}.
\label{eq:triplet_input}
\end{equation}

Each of these triplets is passed through a learnable linear transformation followed by a nonlinearity:

\begin{equation}
o_k^{(i)} = \phi \left( \mathbf{y}_k^{(i)} \cdot \boldsymbol{\theta}^{(i)} + b^{(i)} \right),
\label{eq:triplet_output}
\end{equation}
where \( \boldsymbol{\theta}^{(i)} \in \mathbb{R}^{3 \times F_i} \) denotes the weight matrix of the \(i\)-th fixed kernel layer, \(b^{(i)}\) is the bias vector, \(F_i\) is the number of filters at that layer, and \(\phi(\cdot)\) is the activation function, chosen as \texttt{tanh}. We chose \texttt{tanh} based on both empirical observations---showing no significant performance difference across standard activation functions---and theoretical reasoning: given the deterministic ISI structure shaped by known pulse filters and $\tau$, the network’s reliance on spatial relationships makes it less sensitive to the choice of nonlinearity, with \texttt{tanh} providing sufficient expressiveness for stable and effective learning. The outputs \(o_k^{(i)}\) are concatenated across all \(N\) layers to form a unified feature vector.

\begin{figure*}[ht]
    \centering
    \includegraphics[width=\textwidth]{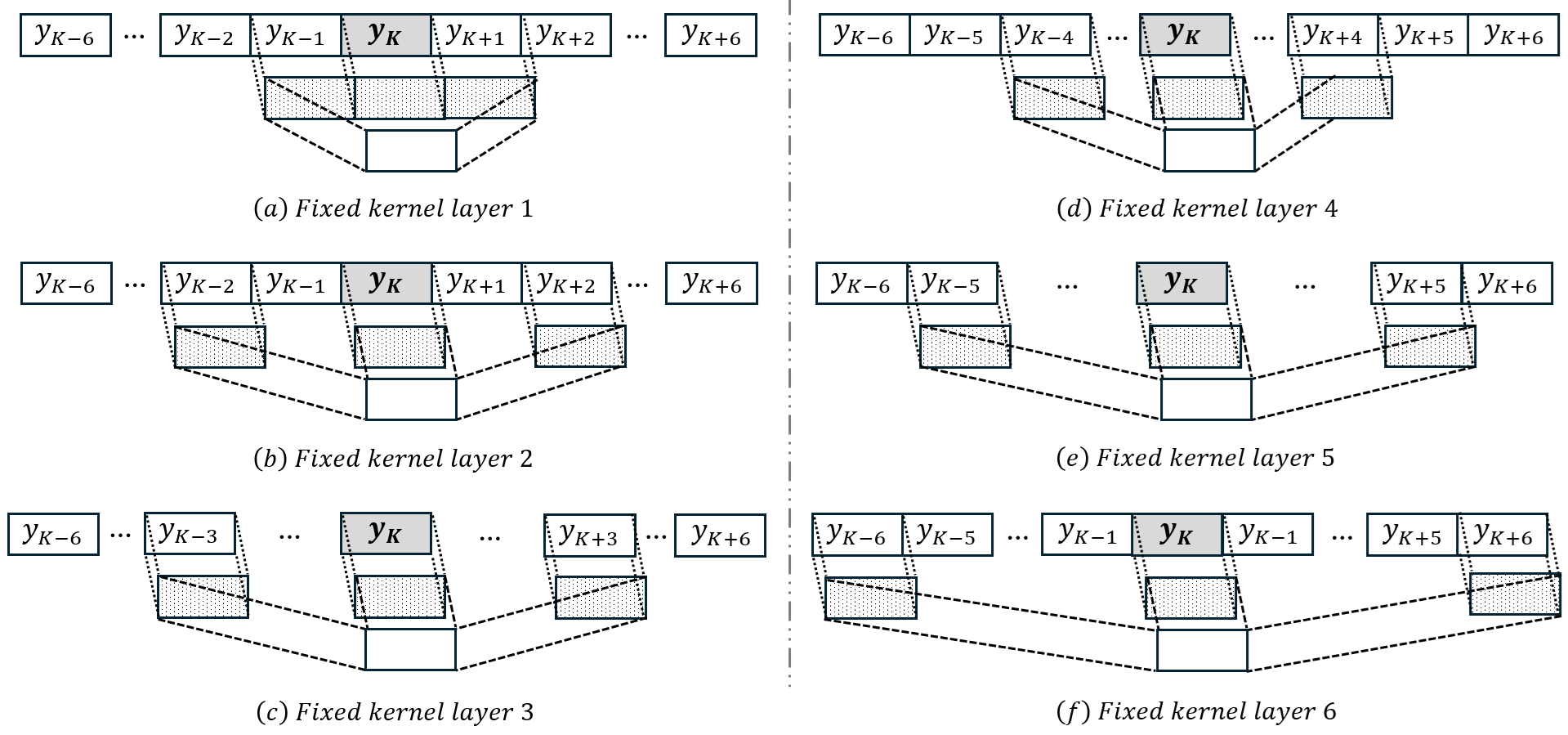}
    \caption{Illustration of structured fixed kernel layers with domain-informed masking: (a) Layer 1 captures symbols one unit from the middle, (b) Layer 2 two units away, etc., up to (f) Layer 6.}
    \label{fig:Fig4_Fixed_Kernel_Layers}
\end{figure*}

This design eliminates the need for sliding kernels and ensures precise alignment with ISI-induced dependencies at each distance \(i\). Domain-informed masking is applied to isolate the corresponding triplets \(\mathbf{y}_k^{(i)}\), allowing the model to focus on specific ISI contributions. For \(N = 6\), the network comprises 6 fixed kernel layers as shown in Fig.~\ref{fig:Fig4_Fixed_Kernel_Layers}. This modular design reflects the symmetric ISI structure described in Section~\ref{sec:system_model}, where interference originates equally from both preceding and succeeding symbols.

Each kernel layer uses a specific number of filters based on the strength of the ISI at the corresponding distance. For instance, symbols closer to \(y_k\) exert stronger interference and are modeled with more filters. As an example, for \(\tau = 0.8\) (with \(N=6\)), the filter allocation is as follows: 4 filters for the first kernel layer, 2 filters for the second and third, and 1 filter each for the remaining layers. The total number of features passed to the subsequent dense layers is thus the sum of filters across all fixed kernels. The configuration is summarized in Table~\ref{tab:filter_numbers}.

\begin{figure}[!t]
\centering
\includegraphics[width=\linewidth]{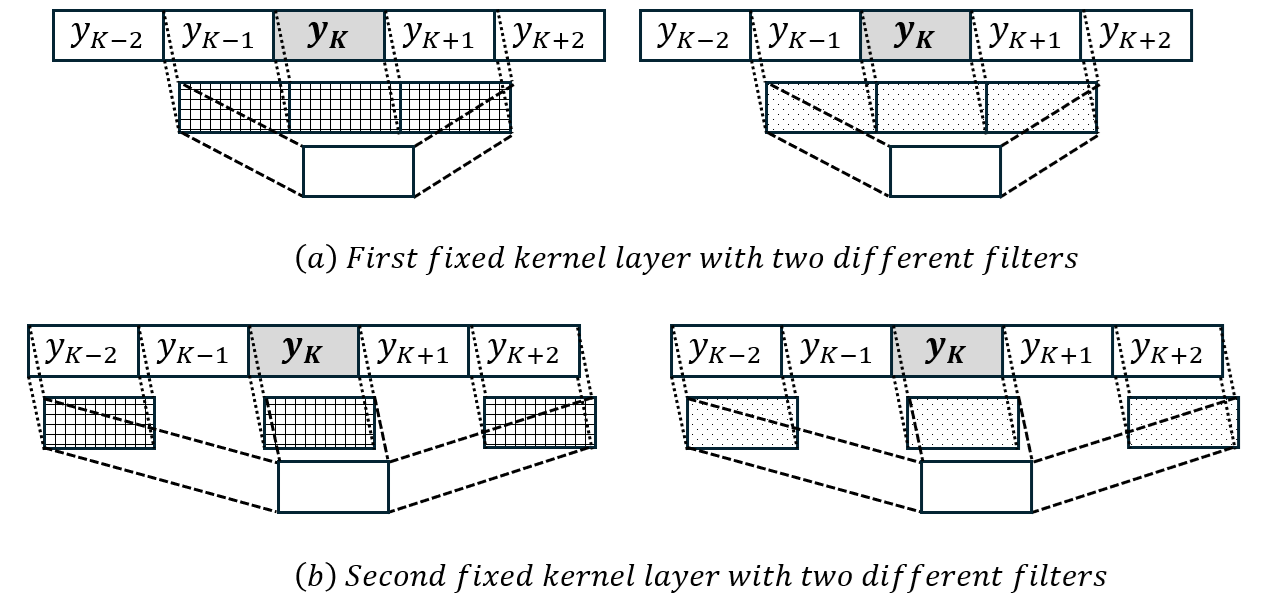}
\caption{Illustration of fixed kernel layers applied with two different filters for \( N=2 \), showing diverse feature extraction.}
\label{fig:Fig5_Filters}
\end{figure}

\begin{table}[ht]
\centering
\caption{Number of filters for each fixed kernel layer under different $\tau$ values in the proposed CNN detector.}
\renewcommand{\arraystretch}{1.3}
\begin{tabular}{lccc}
\rowcolor[gray]{0.9}
 & $N=8$ ($\tau = 0.7$) & $N=6$ ($\tau = 0.8$) & $N=2$ ($\tau = 0.9$) \\
Layer 1 & 8 & 4 & 2 \\
\rowcolor[gray]{0.95}
Layer 2 & 6 & 2 & 1 \\
Layer 3 & 4 & 2 & -- \\
\rowcolor[gray]{0.95}
Layer 4 & 2 & 1 & -- \\
Layer 5 & 2 & 1 & -- \\
\rowcolor[gray]{0.95}
Layer 6 & 1 & 1 & -- \\
Layer 7 & 1 & -- & -- \\
\rowcolor[gray]{0.95}
Layer 8 & 1 & -- & --
\end{tabular}
\label{tab:filter_numbers}
\end{table}

\begin{figure*}[!t]
\centering
\includegraphics[width=\linewidth]{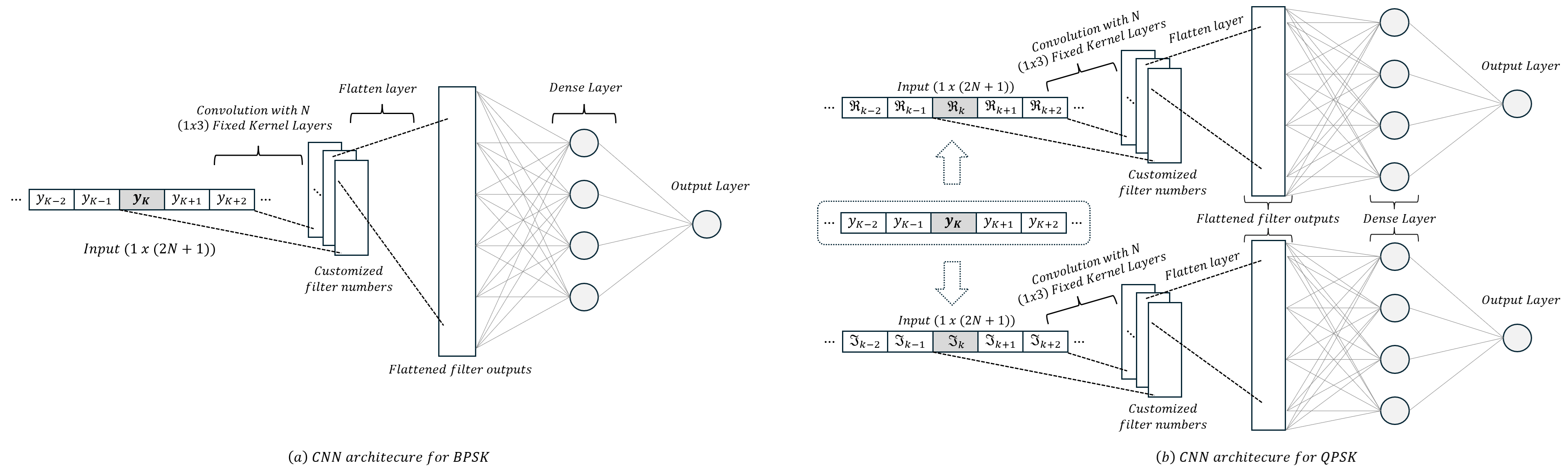}
\caption{Convolutional Neural Network (CNN)-based faster-than-Nyquist (FTN) signaling detector architecture for BPSK and QPSK modulations.}
\label{fig:Fig6_Complete_Cnn_Arch}
\end{figure*}

The feature maps from all fixed kernel layers are concatenated and passed through a fully connected dense layer with 4 neurons. The output of each dense neuron is computed as

\begin{equation}
z_l = \sum_{j=1}^{F_{\text{total}}} \theta^{(l)}_j v_j + b^{(l)}, \quad h_l = \phi(z_l),
\end{equation}
where \(F_{\text{total}}\) is the total number of filters, \(v_j\) is the \(j\)-th feature input, and \(\theta^{(l)}_j\) and \(b^{(l)}\) are the weights and bias of the \(l\)-th neuron. The activation function \(\phi(\cdot)\) is again \texttt{tanh}. The final output layer consists of a single neuron with a sigmoid activation function to produce the probability estimate for the transmitted symbol. The complete CNN-based detector architecture is visualized in Fig.~\ref{fig:Fig6_Complete_Cnn_Arch}.

In the case of complex modulation formats like QPSK, the real and imaginary parts of each received sample are separated and processed independently through identical CNN branches. The output layer then reconstructs the complex-valued symbol decision by combining the two outputs. This design preserves consistency across modulation types while maintaining the network's modularity and efficiency.

\subsection{Complexity Analysis for Structured Fixed Kernel Layers}  
\label{subsec:comp_analysis_FKLs}  

In this section, we analyze the computational efficiency of the proposed structured fixed kernel method compared to the conventional convolutional kernel approach in the context of FTN signaling detection. To provide a concrete illustration, we consider a sequence length of 5 (\(N=2\)), which corresponds to the configuration for \(\tau = 0.9\). As discussed in Section~\ref{subsec:cnn_model}, since the most accurate method among the conventional approaches was found to be the one using a kernel length of 3, as represented in Fig.~\ref{fig:Fig3_Conventional_Kernels}(c), we select the kernel length of 3 as the representative of the conventional kernelization method in our analysis.

\begin{figure}[!t]
\centering
\includegraphics[width=\linewidth]{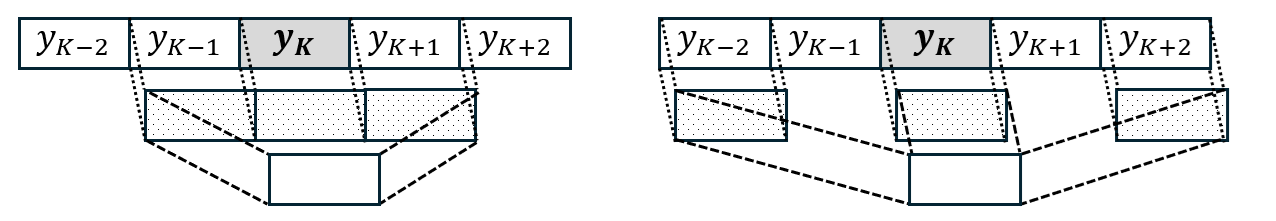}
\caption{Structured fixed kernel layers for a sequence length of 5 (\(N=2\)), designed to capture ISI effects at specific distances from the middle symbol.}
\label{fig:Fig7_FKL_N2}
\end{figure}

In parallel, the structured fixed kernel layers for the same sequence length (\(N=2\)) are illustrated in Fig.~\ref{fig:Fig7_FKL_N2}. Unlike conventional methods, this structured approach explicitly leverages prior knowledge of ISI patterns, minimizes redundant computations, and focuses on learning critical dependencies, thereby optimizing computational efficiency. 

In the conventional kernelization approach, as illustrated in Fig.~\ref{fig:Fig3_Conventional_Kernels}(c), the convolutional kernel moves across the sequence, performing computations at each step to extract features. For a kernel length of 3, the kernel engages in 3 multiplications and 3 additions at each step, including a bias addition, effectively combining the values from the sequence to capture patterns. Given a sequence length of 5 (\(N = 2\)), the kernel requires three steps to traverse the entire sequence. Consequently, the total number of operations amounts to 9 multiplications and 9 additions for a single filter. However, achieving satisfactory performance and precision necessitates the use of at least two filters. This increases the computational cost to 18 multiplications and 18 additions in total. These computations highlight the inherent computational demands of the conventional kernelization method, particularly when multiple filters are employed to enhance precision.

On the other hand, the structured fixed kernel method, depicted in Fig.~\ref{fig:Fig7_FKL_N2}, offers a substantial reduction in computational complexity compared to the conventional kernelization approach. For a sequence of length 5 (\(N = 2\)), this approach requires only two computational steps to extract features, utilizing its domain-informed architecture to eliminate redundancy. Furthermore, as detailed in Table~\ref{tab:filter_numbers}, the first fixed kernel layer employs 2 filters, necessitating a total of 6 multiplications and 6 additions, while the second fixed kernel layer utilizes a single filter, resulting in 3 multiplications and 3 additions. This leads to a cumulative computational cost of 9 multiplications and 9 additions. In contrast, the conventional method requires 18 multiplications and 18 additions under similar conditions, resulting in a 50\% reduction in computational overhead when using the structured fixed kernel method. Importantly, this efficiency gain is achieved without compromising detection performance; rather, the method enhances precision by explicitly modeling ISI effects through specialized fixed kernel layers. A comprehensive analysis of performance in terms of BER will be provided in Section~\ref{subsec:ber_results}.

\subsection{Learning Rate Strategy for Training}
\label{subsec:lr_strategy}

Optimizing the learning rate is a critical aspect of training the proposed CNN-based FTN detector, directly influencing its ability to converge effectively and minimize validation loss over training epochs.While the adaptive moment estimation (Adam) optimizer, renowned for its adaptive learning rate adjustments, serves as the foundation for optimization, its internal mechanisms alone may not suffice in certain scenarios. Specifically, when a constant learning rate is employed, training stagnation can occur, limiting the network's ability to further refine its parameters.

Given these challenges, Adam was selected as the optimizer due to its robustness in handling noisy gradients and its ability to dynamically adjust learning rates for individual parameters. This makes it well-suited for the proposed network’s architecture, which leverages structured fixed kernel layers to address ISI effects at varying distances from the middle symbol. However, despite Adam's adaptive capabilities, relying on a constant learning rate may still lead to suboptimal performance. For instance, in the case of $\tau = 0.8$, validation loss stagnates at approximately $6 \times 10^{-4}$  after the fourth epoch, showing minimal improvement even with extended training.

To overcome this limitation, an exponentially decaying learning rate schedule was introduced alongside Adam. This strategy initializes the learning rate at 0.001 and reduces it gradually by a fixed decay factor after specific intervals during training. This gradual reduction enables the network to transition from coarse updates in earlier epochs to finer weight adjustments in later epochs. By complementing Adam's local adaptability, this global scheduling mechanism helps prevent overshooting and facilitates better convergence.

The impact of this combined approach is evident in the training dynamics. For $\tau = 0.8$, the network employing an exponential decay learning rate achieves significantly better results, with validation loss reducing to approximately $2 \times 10^{-4}$ by the tenth epoch. This improvement underscores the importance of a decaying learning rate in escaping local minima and maintaining consistent progress throughout training. Such a strategy is particularly advantageous in FTN signaling detection, where later epochs demand subtle adjustments to capture intricate ISI patterns effectively. To provide a comprehensive overview of the training configuration, including the learning rate schedule, batch size, optimizer setup, and other training parameters for each $\tau$ value, the complete set of hyperparameters is summarized in Table~\ref{tab:training_hyperparams}.

\begin{table*}[!t]
\centering
\caption{Training Hyperparameters of the Proposed CNN-Based FTN Detector for Different $\tau$ Values}
\label{tab:training_hyperparams}
\renewcommand{\arraystretch}{1.3}
\setlength{\tabcolsep}{8pt}
\begin{tabular}{lccc}
\hline
\rowcolor[gray]{0.9}
\textbf{Parameter} & $\boldsymbol{\tau = 0.9}$ & $\boldsymbol{\tau = 0.8}$ & $\boldsymbol{\tau = 0.7}$ \\
\hline
Number of training samples & 4{,}000{,}000 & 4{,}000{,}000 & 4{,}000{,}000 \\
Number of test samples     & 4{,}000{,}000 & 4{,}000{,}000 & 4{,}000{,}000 \\
SNR values during training & 7, 8, 9, 10 dB & 7, 8, 9, 10 dB & 7, 8, 9, 10 dB \\
Batch size                 & 1000 & 1000 & 1000 \\
Steps per epoch            & 4000 & 4000 & 4000 \\
Number of epochs           & 20 & 20 & 20 \\
Optimizer                  & Adam & Adam & Adam \\
Initial learning rate      & 0.001 & 0.001 & 0.001 \\
Learning rate schedule     & Exponential Decay & Exponential Decay & Exponential Decay \\
Decay steps                & 20{,}000 & 20{,}000 & 20{,}000 \\
Decay interval (epochs)    & Every 5 epochs & Every 5 epochs & Every 5 epochs \\
Decay rate                 & 0.9 & 0.9 & 0.9 \\
Decay staircase            & True & True & True \\
Number of interference symbols ($N$) & 2 & 6 & 8 \\
Modulation formats         & BPSK, QPSK, 16-QAM, 64-QAM & BPSK, QPSK, 16-QAM, 64-QAM & BPSK, QPSK \\
Loss function              & Binary Cross-Entropy & Binary Cross-Entropy & Binary Cross-Entropy \\
Output activation          & Sigmoid & Sigmoid & Sigmoid \\
\hline
\end{tabular}
\end{table*}

\section{Simulation Results and Complexity Comparison}
\label{sec:Simulation Results}

This section presents the simulation results of the proposed CNN-based FTN detector, focusing on both BER performance and computational complexity.

The channel models and simulation conditions are systematically defined as follows. Initially, the signal is transmitted over an AWGN channel, where the noise samples are drawn from a Gaussian distribution $\mathcal{N}(0, N_0/2)$. The SNR is defined as $\text{SNR} = E_s / N_0$, where $E_s$ denotes the average symbol energy after pulse shaping. To evaluate performance under more realistic conditions, a quasi-static multipath Rayleigh fading channel is considered. In this model, the received signal comprises three equal-power Rayleigh-faded paths, with relative delays between adjacent paths set to $\tau T$, where $\tau$ is the FTN compression factor. The fading coefficients $\{h_0, h_1, h_2\}$ are independently drawn from a Rayleigh distribution with unit mean power and remain constant throughout both training and testing phases, representing a quasi-static channel assumption with no Doppler spread. Noise samples are added after multipath propagation to maintain consistency with typical wireless communication models. It is noted that the present study focuses exclusively on wireless channel impairments, and fiber-optic effects such as chromatic dispersion, polarization-mode dispersion, and nonlinearities are not considered.

The BER analysis first compares the structured fixed-kernel approach with conventional CNN architectures under an AWGN channel for \(\tau = 0.9\). This comparison is then extended to more challenging FTN scenarios with \(\tau = 0.8\) and \(\tau = 0.7\), demonstrating the robustness of the proposed detector under severe ISI conditions. Furthermore, to verify the statistical reliability of the reported BER results, 95\% confidence intervals are computed based on binomial statistics, and the corresponding analysis is presented. Beyond AWGN-based evaluations, the detector is assessed over a quasi-static multipath Rayleigh fading channel, where its BER performance is benchmarked against the optimal BCJR algorithm across multiple compression factors. In addition, a spectral efficiency analysis is performed, quantifying the achievable data rate as a function of the compression factor and modulation format under FTN signaling. To further validate the practicality of the proposed method, a comprehensive computational complexity analysis is conducted, including operation-level comparisons and hardware resource evaluations relative to the M-BCJR algorithm and other existing FTN detection approaches.

\subsection{BER Results}
\label{subsec:ber_results}
\begin{figure}[!t]
\centering
\includegraphics[width=\linewidth]{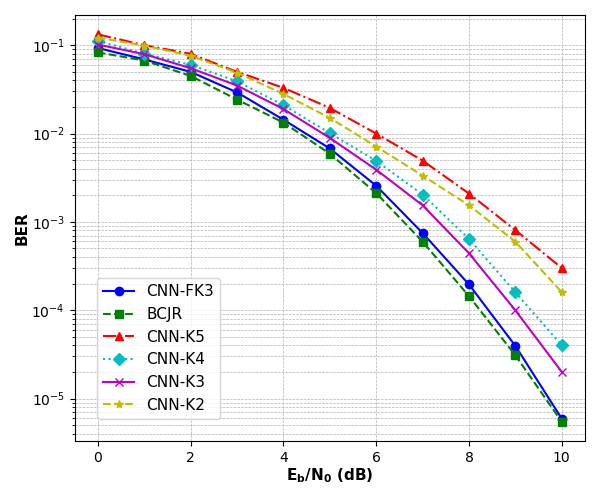}
\caption{Comparison of the BER performance of the proposed CNN-based FTN detector using structured fixed kernel layers (CNN-FK3) with conventional kernel-based CNN architectures (CNN-K2, CNN-K3, CNN-K4, CNN-K5) for $\tau = 0.9$, with a roll-off factor of $\beta = 0.35$, under an AWGN channel.}
\label{fig:Fig8_Fixed_vs_Conventional}
\end{figure}

In order to evaluate the effectiveness of the proposed CNN-based FTN signaling detector employing structured fixed kernel layers with domain-informed masking, we firstly compared it with conventional kernel-based CNN architectures under the configuration of $\tau = 0.9$, where the sequence length is 5 ($N = 2$). This comparison was specifically conducted for $\tau = 0.9$ because conventional kernel methods struggle to achieve performance comparable to the optimal BCJR algorithm at lower $\tau$ values, such as $\tau = 0.8$ and $\tau = 0.7$, due to their limited ability to effectively capture the increased ISI effects in these scenarios. The results of this comparison are depicted in Fig.~\ref{fig:Fig8_Fixed_vs_Conventional}. In the coming figures, ``CNN-FK3'' represents the proposed CNN detector utilizing the structured fixed kernel approach, where ``FK'' denotes fixed kernel and ``3'' reflects the kernel length. As described in Section~\ref{subsec:cnn_model}, this kernel configuration extracts data from the middle symbol and its nearest neighbors at equal distances, achieving a kernel length of 3. In contrast, ``CNN-K2,'' ``CNN-K3,'' ``CNN-K4,'' and ``CNN-K5'' represent CNN architectures using conventional kernels of lengths 2, 3, 4 and 5, respectively. The structures of this kernels were given in Fig.~\ref{fig:Fig3_Conventional_Kernels}.

\begin{figure}[!t]
\centering
\includegraphics[width=\linewidth]{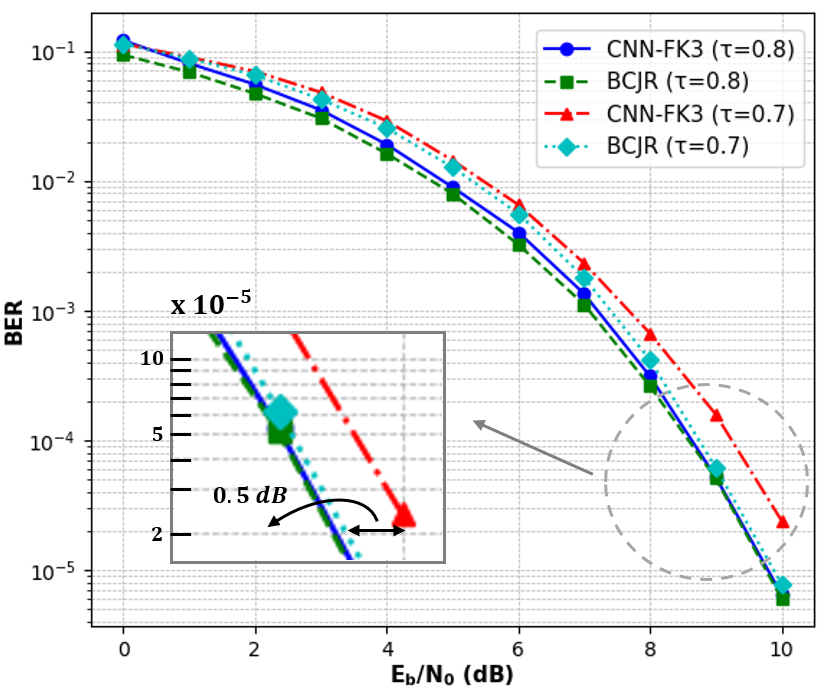}
\caption{Comparison of the BER performance of the proposed CNN-based FTN detector with the optimal BCJR algorithm for $\tau = 0.8$ and $\tau = 0.7$, with a roll-off factor of $\beta = 0.35$, under an AWGN channel.}
\label{fig:Fig9_Cnnfk_vs_Bcjr}
\end{figure}

\begin{figure}[!t]
\centering
\includegraphics[width=\linewidth]{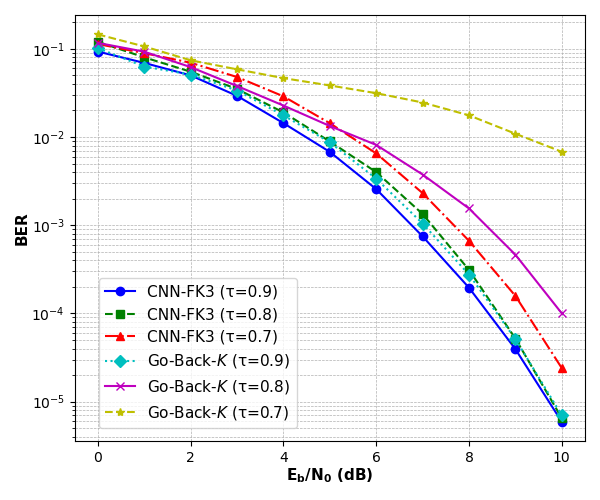}
\caption{BER performance comparison between the proposed CNN-based FTN detector and the Go-Back-\textit{K} algorithm [5] for different compression factors (\(\tau = 0.9\), \(\tau = 0.8\), and \(\tau = 0.7\)), with a roll-off factor of $\beta = 0.35$, under an AWGN channel.}
\label{fig:Fig10_Cnnfk_vs_GobackK}
\end{figure}

\begin{figure}[!t]
\centering
\includegraphics[width=\linewidth]{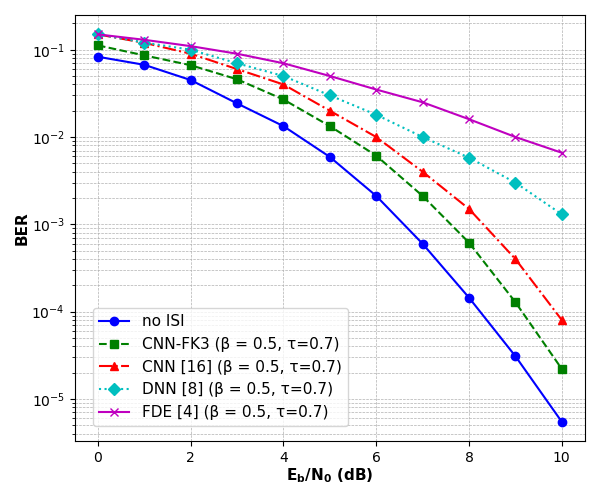}
\caption{Comparison of the BER performance of the proposed CNN-based FTN detector with the methods in \cite{sugiura2013}, \cite{song2020}, and \cite{defilippo2025} for $\tau = 0.7$ and $\beta = 0.5$, under an AWGN channel.}
\label{fig:Fig11_Cnnfk_vs_Others}
\end{figure}

The results in Fig.~\ref{fig:Fig8_Fixed_vs_Conventional} clearly demonstrate the superiority of the structured fixed kernel approach. CNN-FK3 aligns closely with the optimal BCJR algorithm, achieving nearly identical BER performance for $\tau = 0.9$. Conventional kernel methods, on the other hand, exhibit significantly inferior results, with CNN-K3 showing the best performance among them. This outcome can be explained by the limitations of conventional kernels in effectively capturing ISI dynamics. Larger kernels, such as a kernel size of 5, depicted in Fig.~\ref{fig:Fig3_Conventional_Kernels}(a), provide a broader receptive field, diluting the localized ISI effects. In contrast, smaller kernels, like a kernel size of 2, represented in Fig.~\ref{fig:Fig3_Conventional_Kernels}(d), suffer from a limited receptive field, restricting the network's ability to capture sufficient contextual information about ISI. Kernel sizes of 4 and 3, shown in Fig.~\ref{fig:Fig3_Conventional_Kernels}(b) and Fig.~\ref{fig:Fig3_Conventional_Kernels}(c), perform significantly better. A kernel size of 3 is particularly effective as it strikes a balance between capturing ISI effects and maintaining alignment with the middle symbol, as discussed in Section~\ref{subsec:cnn_model}. However, despite its relative effectiveness, it still falls short of the performance achieved by the structured fixed kernel approach, which is specifically designed to target ISI effects with greater precision and efficiency. The structured fixed kernel method outperforms conventional approaches by addressing the inherent limitations of moving kernels and leveraging domain knowledge to focus on localized ISI effects. This approach not only achieves superior performance but also aligns with the computational efficiency considerations discussed in Section~\ref{subsec:comp_analysis_FKLs}. 

To further evaluate the performance of the proposed CNN-based FTN detector, experiments were conducted at lower $\tau$ values, specifically $\tau = 0.8$ and $\tau = 0.7$. These scenarios represent increased ISI effects, posing a greater challenge for detection algorithms. The results are presented in Fig.~\ref{fig:Fig9_Cnnfk_vs_Bcjr}, which demonstrates the robustness of the proposed method under these challenging conditions. Since the BER performance for BPSK and QPSK modulation schemes is nearly identical, a single graph is provided to represent both cases, ensuring clarity and conciseness in the presentation.

For $\tau = 0.8$, the proposed detector aligns remarkably well with the optimal BCJR performance across all SNR values, achieving near-identical BER results. This underscores the efficacy of the structured fixed kernel approach in capturing and mitigating ISI effects.

For $\tau = 0.7$, where ISI effects are more pronounced, the proposed method maintains strong performance, with only a slight deviation of approximately 0.5 dB from the BCJR algorithm at higher SNR values for a BER of $2 \times 10^{-5}$. While this gap might seem like a limitation, achieving such performance with a CNN-based architecture is a significant accomplishment, as CNNs are often combined with other architectures, such as Bi-LSTM, in FTN signaling \cite{yang2024mhsa}, \cite{liu2022datadriven}. The proposed method challenges this convention by demonstrating that structured fixed kernels enable CNNs to effectively perform detection as a standalone solution. Furthermore, the detector’s robustness extends across various $\tau$ values, including lower values such as $\tau = 0.7$, highlighting its adaptability and potential as an efficient alternative to traditional FTN detection algorithms.

To evaluate the proposed CNN-based FTN detector's ability to achieve high performance with lower complexity, we compare its BER performance with the low-complexity successive symbol-by-symbol sequence estimator presented in \cite{bedeer2017}, considering a roll-off factor of $\beta = 0.35$. Since the method in \cite{bedeer2017} also focuses on low-complexity detection, it serves as a suitable benchmark for assessing both the efficiency and accuracy of the proposed approach. The BER comparison is illustrated in Fig.~\ref{fig:Fig10_Cnnfk_vs_GobackK}. For \(\tau = 0.9\), the proposed CNN-based detector and the method in \cite{bedeer2017} deliver nearly identical BER results, reflecting comparable effectiveness under conditions of mild ISI. However, as the compression factor decreases to \(\tau = 0.8\) and \(\tau = 0.7\), the superiority of the proposed method becomes evident. Specifically, the proposed method achieves better BER performance, effectively handling the pronounced ISI effects associated with lower \(\tau\) values. In addition to BER performance, these two algorithms will also be compared in terms of computational complexity in Section~\ref{subsection:last_comp}. This dual comparison provides a comprehensive evaluation of the proposed method, emphasizing both its accuracy and efficiency. By demonstrating superior performance over another low-complexity approach, the proposed method establishes itself as an efficient and accurate alternative for FTN signaling detection, further validating its potential for practical applications.

To further assess the effectiveness of the proposed CNN-based FTN detector, we compare its BER performance with the methods proposed in \cite{sugiura2013}, \cite{song2020}, and \cite{defilippo2025}. For consistency, all methods are evaluated at $\tau = 0.7$ and $\beta = 0.5$. As shown in Fig.~\ref{fig:Fig11_Cnnfk_vs_Others}, the proposed method achieves a lower BER across a wide range of SNR values while also demonstrating improved computational efficiency, which is further discussed in Section~\ref{subsection:last_comp}.

\begin{figure}[!t]
\centering
\includegraphics[width=\linewidth]{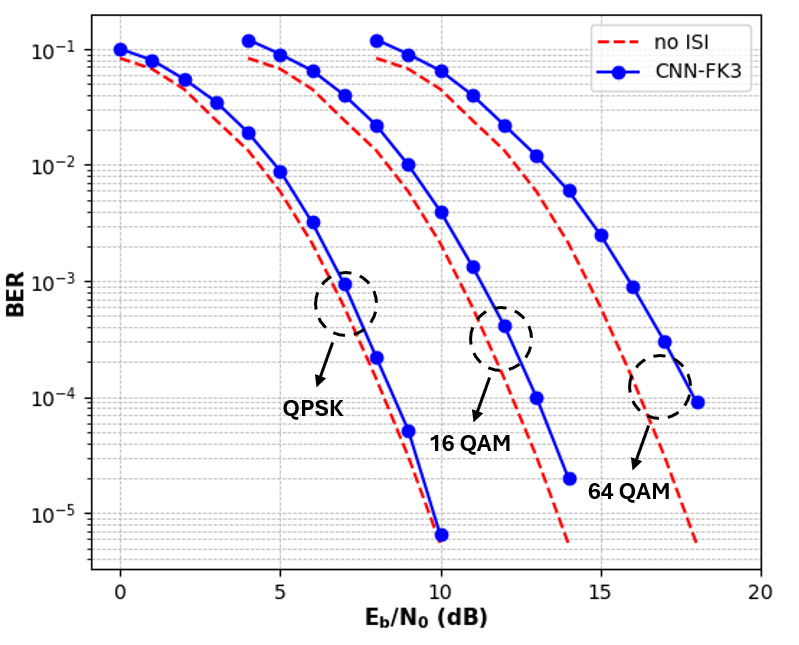}
\caption{BER performance of the proposed CNN-based FTN detector for QPSK, 16-QAM and 64-QAM, with $\tau = 0.8$ and $\beta = 0.5$, under an AWGN channel.}
\label{fig:Fig12_64Qam}
\end{figure}

\begin{figure}[!t]
\centering
\includegraphics[width=\linewidth]{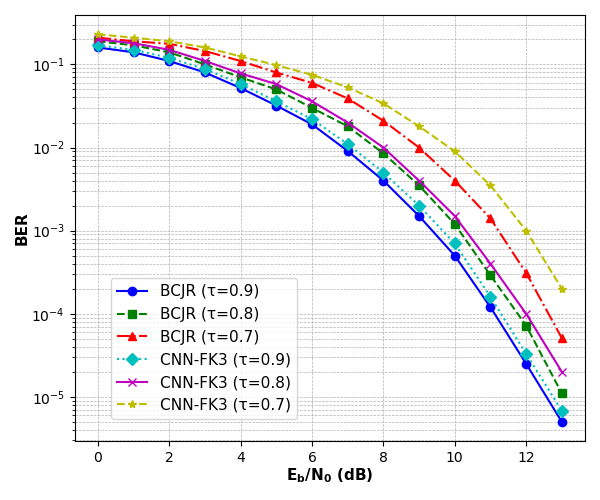}
\caption{BER performance of the proposed CNN-FK3 detector under quasi-static multipath Rayleigh fading channel for $\tau = 0.7$, $\tau = 0.8$, and $\tau = 0.9$ with QPSK modulation. The benchmark BCJR results are also included for comparison.}
\label{fig:Fig13_FadingChannel}
\end{figure}

In addition to QPSK, the proposed method is evaluated for higher-order modulations, specifically 16-QAM and 64-QAM, using the same network topology suggested for QPSK in Fig.~\ref{fig:Fig6_Complete_Cnn_Arch}(b). The simulation is conducted for $\tau = 0.8$ and $\beta = 0.5$, and the results are presented in Fig.~~\ref{fig:Fig12_64Qam}, demonstrating the method's adaptability to higher modulation orders.

To evaluate the detector’s robustness under more realistic wireless environments, an additional set of experiments was conducted over a quasi-static multipath fading channel. In this scenario, a Rayleigh fading model with three equal-power paths was considered. The relative delay between adjacent multipath components was set equal to the symbol interval \(\tau T\), causing each delayed replica to align with a subsequent symbol position and thereby intensifying the ISI effect. To emulate a quasi-static fading condition, the channel coefficients were assumed to remain constant within each frame and to change independently across different frames during both training and testing phases. The proposed CNN-based FTN detector was compared with the optimal BCJR algorithm for \(\tau = 0.9\), \(\tau = 0.8\), and \(\tau = 0.7\), using QPSK modulation. The results, shown in Fig.~\ref{fig:Fig13_FadingChannel}, demonstrate that the proposed method maintains strong detection performance even in the presence of multipath fading. Notably, it achieves BER performance closely matching that of the BCJR algorithm across all \(\tau\) values, with only minor deviations observed at \(\tau = 0.7\). This near-optimal performance in a fading environment further confirms the effectiveness of the structured fixed kernel design in learning and compensating for channel-induced distortions, validating the proposed method’s applicability in practical FTN systems beyond AWGN-only scenarios.

To assess the performance of the proposed CNN-based FTN detector under practical communication settings, we incorporate forward error correction using LDPC codes. LDPC codes are well-established in modern communication standards such as 5G NR and IEEE 802.11, and are known for their capacity-approaching performance with iterative decoding. In this work, we employ a standardized LDPC code with a block length of $n = 1056$ bits and a code rate of $R = 1/2$, corresponding to $k = 528$ information bits and $528$ parity bits. The parity-check matrix used in the simulations is derived from the base graph structures defined in the 3GPP 5G NR standard (3GPP TS 38.212), ensuring compatibility with real-world deployment scenarios and providing reliable decoding performance.

At the transmitter side, the input bitstream is encoded using the LDPC encoder, and the resulting 1056-bit codeword is modulated using BPSK, where bits $\{0, 1\}$ are mapped to symbols $\{+1, -1\}$. The modulated symbols are then passed through a RRC filter and transmitted using FTN signaling with three different time packing factors: $\tau = 0.9$, $\tau = 0.8$, and $\tau = 0.7$. The received signal, affected by ISI due to time compression and AWGN, is matched-filtered and sampled at the receiver at intervals of $\tau T$.

The sampled signal is processed by the proposed CNN-based detector, which is trained to exploit the structured ISI patterns introduced by FTN signaling. For each target bit, a window of $(2N+1)$ adjacent samples is fed into the CNN, which outputs a real-valued soft estimate corresponding to the likelihood of the central symbol. These outputs are mapped to log-likelihood ratios (LLRs) and serve as input to the LDPC decoder. The decoder operates using the belief propagation (BP) algorithm with a maximum of 50 iterations.

\begin{figure}[!t]
\centering
\includegraphics[width=\linewidth]{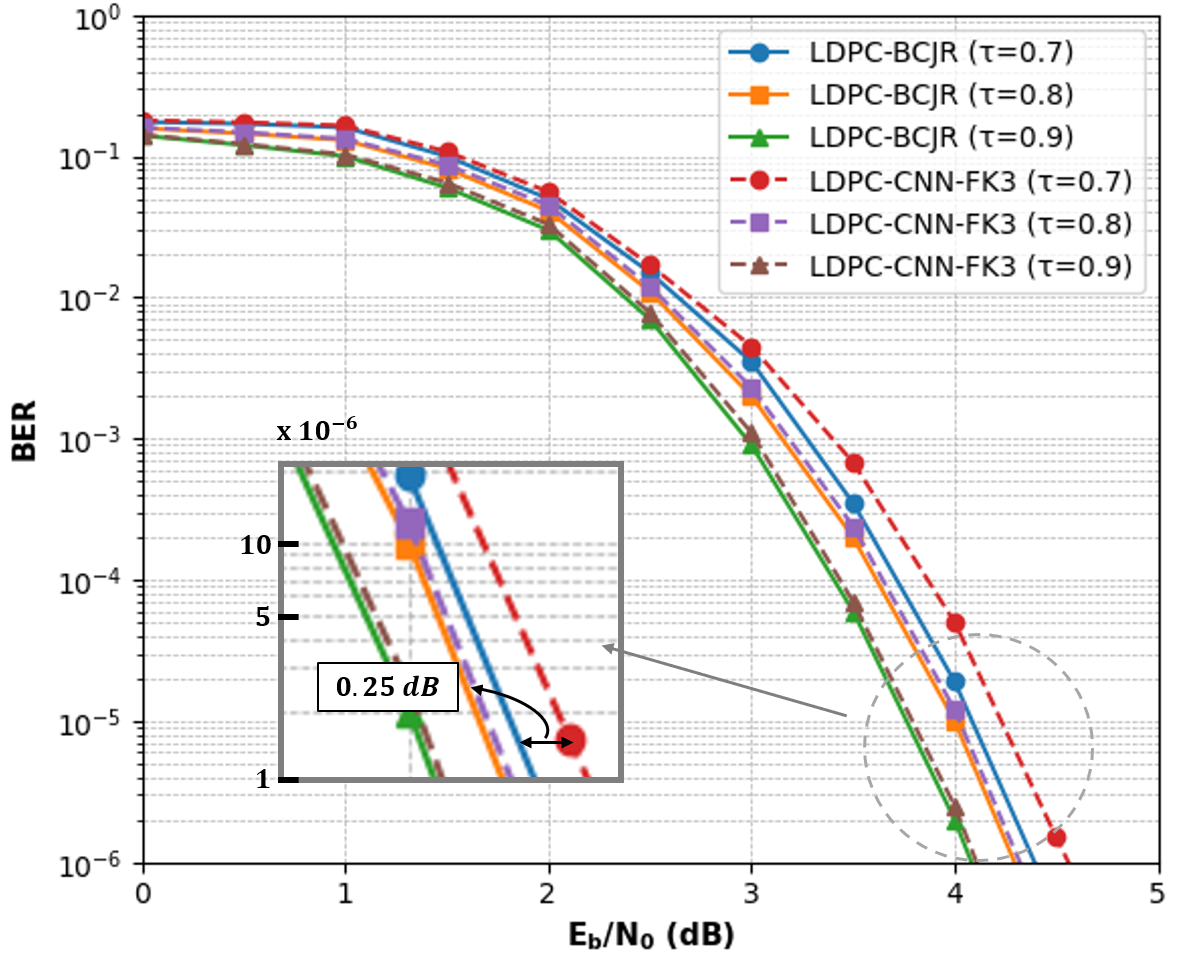}
\caption{Bit error rate (BER) performance comparison of LDPC-coded FTN systems using BCJR-based and CNN-based detectors for $\tau = 0.7$, $\tau = 0.8$, and $\tau = 0.9$. The CNN-based detector employs a fixed kernel (FK3) architecture. All configurations use a rate-1/2 LDPC code of length 1056 bits under BPSK modulation over an AWGN channel.}
\label{fig:Fig14_LDPC}
\end{figure}

To benchmark the performance of the LDPC-coded system, we compare the results with those obtained using the optimal BCJR algorithm under the same FTN signaling conditions. The BER performance of both systems is evaluated across a range of SNRs, and the results are presented in Fig.~\ref{fig:Fig14_LDPC}. The inclusion of LDPC coding significantly improves the detection robustness, especially in the presence of strong ISI for lower $\tau$ values. Moreover, the CNN-based detector, when combined with LDPC decoding, achieves performance levels that are comparable to the BCJR algorithm, while offering substantial reductions in computational complexity. These results demonstrate the practicality and efficiency of the proposed deep learning-based detection framework in coded FTN communication systems.

To complement the BER-based performance assessment, a spectral efficiency analysis was conducted to illustrate the trade-off between reliability and bandwidth utilization across different modulation orders and FTN compression factors. The spectral efficiency (SE), defined as the number of bits transmitted per second per Hertz of bandwidth, is mathematically expressed as
\begin{equation}
\text{SE} = \frac{\log_2 M}{\tau (1 + \beta)} \quad \text{[bps/Hz]},
\end{equation}
where $M$ denotes the modulation order, $\tau$ is the FTN time compression factor, and $\beta$ is the roll-off factor of the employed pulse shaping filter. This expression accounts for both the modulation efficiency via $\log_2 M$ and the bandwidth expansion caused by excess bandwidth $\beta$ in the RRC filter. In our simulations, the roll-off factor $\beta$ was set to $0.35$ for BPSK and QPSK modulations, and $0.5$ for 16-QAM and 64-QAM, aligning with commonly adopted values in the literature.

\begin{figure}[!t]
\centering
\includegraphics[width=\linewidth]{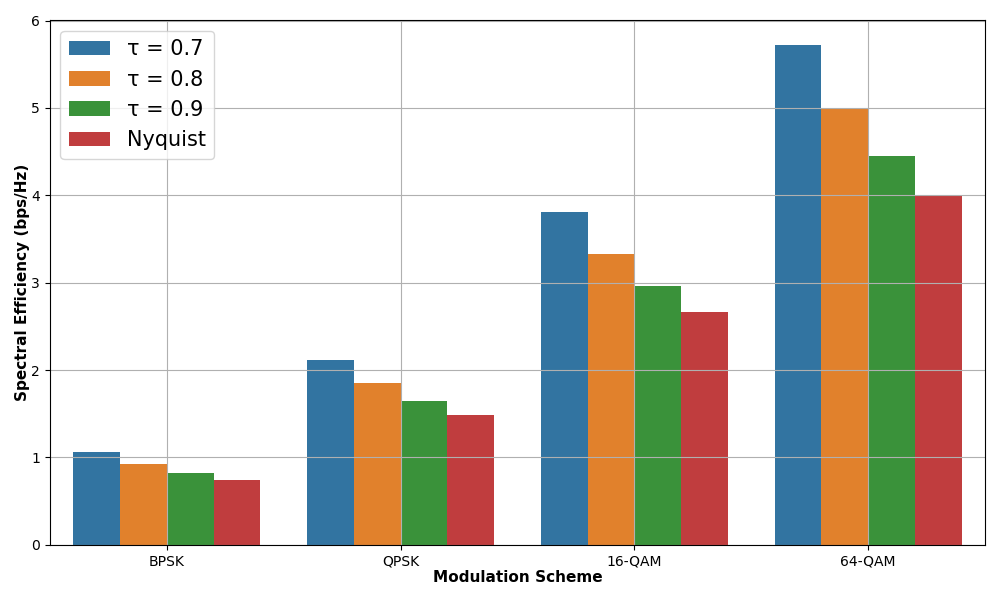}
\caption{Spectral efficiency comparisons for BPSK, QPSK, 16-QAM, and 64-QAM under FTN configurations with $\tau = 0.7$, $\tau = 0.8$, $\tau = 0.9$, and the Nyquist case ($\tau = 1$). Roll-off factor $\beta = 0.35$ is used for BPSK/QPSK and $\beta = 0.5$ for 16-QAM/64-QAM.}
\label{fig:Fig15_Spectral_eff}
\end{figure}

Fig.~\ref{fig:Fig15_Spectral_eff} presents the resulting spectral efficiencies for $\tau = 0.7$, $\tau = 0.8$, $\tau = 0.9$, and the Nyquist case ($\tau = 1$), offering a comparative perspective across four modulation formats. Due to BER degradation at lower $\tau$ values in higher-order modulations, the evaluations for 16-QAM and 64-QAM are limited to $\tau \geq 0.8$ to ensure practical detectability. As expected, spectral efficiency increases with decreasing $\tau$, highlighting the fundamental benefit of FTN signaling in terms of bandwidth compression. These values provide a quantitative reference for balancing system throughput and complexity, reinforcing the proposed detector's viability in spectrally constrained environments.

\subsection{Computational Complexity Comparison}
\label{subsection:last_comp}
A vital consideration when assessing any detection algorithm is its computational efficiency, especially in environments with limited resources. To address this, we conducted an in-depth analysis to compare the inference computational complexity of the proposed CNN-based FTN detector with the traditional M-BCJR algorithm. This analysis evaluates the number of operations required to detect 100 symbols modulated using BPSK and QPSK, across various values of the parameter $\tau$ ($\tau = 0.7$, $0.8$, and $0.9$). 

To provide the reader with insight into the determination of operation numbers for the CNN-based FTN detector, we focus on the case of BPSK-modulated signal detection for \(\tau = 0.8\). For the sake of brevity, the detailed computational analysis for \(\tau = 0.7\) and \(\tau = 0.9\) is omitted.

The operations performed within each fixed kernel layer are consistent with the structure defined in Section~\ref{subsec:cnn_model}. For a given ISI distance \(i \in \{1, 2, \dots, N\}\), the input triplet centered at symbol index \(k\) is defined in~(\ref{eq:triplet_input}). Each of these triplets is processed by a weight matrix \( \boldsymbol{\theta}^{(i)} \in \mathbb{R}^{3 \times F_i} \) and a bias vector \( b^{(i)} \in \mathbb{R}^{F_i} \), followed by an element-wise nonlinearity as described in~(\ref{eq:triplet_output}). All outputs across \(N\) fixed kernel layers are concatenated to form the unified feature vector for the current symbol.

For $\tau = 0.8$, as indicated in Table~\ref{tab:filter_numbers}, the proposed CNN-based detector employs the following filter configuration across the fixed kernel layers: the first fixed kernel layer uses 4 filters, the second and third layers each use 2 filters, while the fourth, fifth, and sixth layers each use 1 filter. For each kernel operation, feature extraction involves 3 multiplications, 3 additions, and 1 \texttt{tanh} operation. With a total of 11 filters, the fixed kernel layers collectively require 33 multiplications, 33 additions, and 11 \texttt{tanh} operations. 

The computation for each neuron in the dense layer follows the formulation introduced in Section~\ref{subsec:cnn_model}. Let \(F_{\text{total}}\) denote the total number of filters used across all fixed kernel layers, and let \(\mathbf{v} \in \mathbb{R}^{F_{\text{total}}}\) represent the concatenated feature vector obtained after flattening the outputs of the fixed kernel layers. Then, the pre-activation value for the \(l\)-th neuron is given as
\begin{equation}
z_l = \sum_{j=1}^{F_{\text{total}}} \theta^{(l)}_j v_j + b^{(l)},
\end{equation}
where \(\theta^{(l)}_j\) and \(b^{(l)}\) denote the weight and bias parameters associated with the \(l\)-th neuron, respectively. The activated output is computed as
\begin{equation}
h_l = \phi(z_l),
\end{equation}
where \(\phi(\cdot)\) denotes the \texttt{tanh} activation function. This process is repeated for all four neurons in the dense layer. The final layer consists of a single neuron with a sigmoid activation function to yield a probability estimate for the transmitted symbol. The final output \( \hat{a}_k \) of the network is obtained by applying a sigmoid activation function to the output of the dense layer. Specifically, for the final output neuron, the computation is expressed as
\begin{equation}
\hat{a}_k = \sigma\left( \sum_{j=1}^{F_{\text{dense}}} \theta^{\text{(out)}}_j h_j + b^{\text{(out)}} \right),
\end{equation}
where \(F_{\text{dense}} = 4\) is the number of neurons in the preceding dense layer, \(h_j\) are the outputs from the dense layer, \(\theta^{\text{(out)}}_j\) and \(b^{\text{(out)}}\) denote the weights and bias for the output neuron, and \(\sigma(\cdot)\) represents the sigmoid activation function. This operation yields the probability estimate corresponding to the central transmitted symbol. In the proposed method, the  \texttt{tanh} activation function is employed for the dense layer with 4 neurons, while the  \texttt{sigmoid} function is applied in the output layer. Consequently, the dense layer involves \(4 \times 11 = 44\) multiplications, \(4 \times 11 = 44\) additions for the bias terms, \(4 \times 10 = 40\) additions resulting from the summation in the activation function, and \(4 \times 1 = 4\) \texttt{tanh} operations. 

Finally, the output layer, which comprises a single neuron, performs 4 multiplications, 4 additions for the bias term, 3 additions resulting from the summation in the activation function, and 1 \texttt{sigmoid} operation. 

Therefore, the CNN-based FTN detector requires a total of 81 multiplications, 124 additions, 15 \texttt{tanh} operations, and 1 \texttt{sigmoid} operation to detect one BPSK-modulated symbol for \(\tau = 0.8\). As illustrated in the proposed CNN architecture in Fig.~\ref{fig:Fig6_Complete_Cnn_Arch}(b), these operation numbers will simply be doubled for the detection of QPSK-modulated symbols. Consequently, for the detection of 1 QPSK-modulated symbols, the proposed method requires 162 multiplications, 248 additions, 30 \texttt{tanh} operations, and 2 \texttt{sigmoid} operations for \(\tau = 0.8\).

Finally, Table~\ref{tab:comp_bpsk_qpsk} presents a comprehensive summary of the operation counts for both the M-BCJR algorithm \cite{kokshoorn2016} and the proposed CNN-based detector across \(\tau = 0.7\), \(\tau = 0.8\), and \(\tau = 0.9\), detailing the computational requirements for detecting 100 symbols with BPSK- and QPSK-modulated signals.

\begin{table}[h!]
\centering
\caption{Comparison of Operation Numbers for M-BCJR and CNN-Based FTN Detector (100 Symbols)}
\resizebox{\columnwidth}{!}{%
\begin{tabular}{@{}llcccccc@{}}
\toprule
\textbf{Operation} & \textbf{Algorithm} & \multicolumn{3}{c}{\textbf{BPSK Modulation}} & \multicolumn{3}{c}{\textbf{QPSK Modulation}} \\
\cmidrule(lr){3-5} \cmidrule(lr){6-8}
& & \textbf{\(\tau=0.7\)} & \textbf{\(\tau=0.8\)} & \textbf{\(\tau=0.9\)} & \textbf{\(\tau=0.7\)} & \textbf{\(\tau=0.8\)} & \textbf{\(\tau=0.9\)} \\
\midrule
Multiplication & M-BCJR & 4631 & 2783 & 1877 & 33048 & 19964 & 13916 \\
               & CNN    & 17900 & 8100 & 2500 & 35800 & 16200 & 5000 \\
\midrule
Addition       & M-BCJR & 9982 & 5129 & 2584 & 70857 & 36137 & 18671 \\
               & CNN    & 27800 & 12400 & 3600 & 55600 & 24800 & 7200 \\
\midrule
Comparison     & M-BCJR & 4160 & 1056 & 272 & 29283 & 7439 & 1976 \\
               & CNN    & --   & --   & --  & --   & --   & -- \\
\midrule
Division       & M-BCJR & 3982 & 2981 & 1436 & 28819 & 21857 & 9488 \\
               & CNN    & --   & --   & --  & --   & --   & -- \\
\midrule
Exponential    & M-BCJR & 205  & 204  & 203  & 410   & 408   & 406 \\
               & CNN    & --   & --   & --  & --   & --   & -- \\
\midrule
Tanh           & M-BCJR & 104  & 103  & 102  & 208   & 206   & 204 \\
               & CNN    & 2900 & 1500 & 700  & 5800  & 3000  & 1400 \\
\midrule
Sigmoid        & M-BCJR & --   & --   & --  & --   & --   & -- \\
               & CNN    & 100  & 100  & 100  & 200  & 200  & 200 \\
\bottomrule
\end{tabular}%
}
\label{tab:comp_bpsk_qpsk}
\end{table}

As the CNN-based detector and the M-BCJR algorithm rely on fundamentally different types of operations, directly comparing their computational demands solely based on the number of operations would be inadequate. To ensure a fair and meaningful evaluation, we utilized a widely accepted FPGA resource estimation tool to determine the hardware resource requirements of each operation, expressed in terms of LUTs. Since LUT utilization provides a standardized measure of hardware complexity across different types of operations, it enables a more objective comparison of the computational efficiency of both approaches. It is worth noting that LUT utilization not only reflects the hardware resource consumption but also serves as an indicative metric for dynamic energy efficiency in FPGA-based implementations. Since LUTs constitute a major portion of the switching logic elements, a lower LUT usage typically correlates with reduced switching activity and consequently lower dynamic power consumption \cite{lut_energy_fpga}. Therefore, the LUT-weighted computational cost metric used in this work provides a meaningful insight into both hardware efficiency and relative energy consumption across the compared methods. These estimations were conducted assuming 10-bit numerical precision, with the resulting resource utilizations detailed in Table~\ref{tab:lut_util}.

\begin{table}[h!]
\centering
\caption{LUT Utilizations for Different Operations (10-bit)}
\resizebox{\columnwidth}{!}{%
\begin{tabular}{@{}lccccccc@{}}
\toprule
\textbf{Operation} & \textbf{Multiplication} & \textbf{Addition} & \textbf{Comparison} & \textbf{Division} & \textbf{Exponential} & \textbf{tanh} & \textbf{sigmoid} \\
\midrule
LUT Utilization & 113 & 10 & 11 & 236 & 73 & 1 & 1 \\
\bottomrule
\end{tabular}%
}
\label{tab:lut_util}
\end{table}

Using the operational data provided in Table~\ref{tab:comp_bpsk_qpsk}, combined with the LUT-based hardware requirements detailed in Table~\ref{tab:lut_util}, a comprehensive metric was devised to assess the overall computational complexity of the two algorithms. This metric, obtained by multiplying the number of operations by their corresponding resource utilization values, serves as an integrated measure of computational cost. By accounting for both the quantity and type of operations alongside their hardware resource demands, this evaluation offers a holistic perspective on computational efficiency. 

\begin{table}[h!]
\centering
\caption{LUT-Weighted Computational Cost Metrics for M-BCJR and CNN-Based FTN Detector (BPSK and QPSK Modulations)}
\begin{tabular}{@{}lccccc@{}}
\toprule
\textbf{Modulation} & \textbf{Algorithm} & \textbf{\(\tau=0.7\)} & \textbf{\(\tau=0.8\)} & \textbf{\(\tau=0.9\)} \\
\midrule
BPSK & M-BCJR & 1623704 & 1095896 & 594750 \\
     & CNN    & 2303700 & 1040900  & 319300 \\
\midrule
QPSK & M-BCJR & 11596529& 7887373 & 4049964\\
     & CNN    & 4607400 & 2081800 & 638600 \\
\bottomrule
\end{tabular}
\label{tab:total_comp}
\end{table}

The results of the computational efficiency analysis, summarized in Table~\ref{tab:total_comp}, reveal the significant advantages of the proposed CNN-based FTN detector, particularly for QPSK modulation. For \(\tau = 0.9\) and \(\tau = 0.8\), the proposed method achieves notable computational efficiency improvements over the M-BCJR algorithm, with reductions of 46\% and 5\%, respectively, for BPSK modulation, as shown in Table~\ref{tab:comp_percent}. Furthermore, for QPSK modulation, the reductions are even more pronounced, reaching 84\% for \(\tau = 0.9\), 74\% for \(\tau = 0.8\), and 60\% for \(\tau = 0.7\). This disparity arises from the distinct growth patterns in computational complexity between the two methods. Specifically, while the computational complexity of the M-BCJR algorithm scales exponentially with the modulation order, the CNN-based detector exhibits linear growth, making it particularly well-suited for higher-order modulation schemes such as QPSK. This reduction stems from the fundamental algorithmic structures of the two methods rather than specific implementations, as low-level code optimizations (e.g., rewriting expressions to reduce arithmetic operations) do not affect asymptotic complexity. In particular, our analysis, grounded in a careful examination of the implementation to ensure fidelity to the original algorithm, confirms that the CNN-based approach achieves up to 84\% theoretical complexity reduction over the M-BCJR algorithm by leveraging parallel processing and learned weights, independent of software or hardware optimizations.

\begin{table}[h!]
\centering
\caption{Percentage Improvement of CNN over M-BCJR in LUT-Weighted Computational Cost Metric}
\label{tab:comp_percent}
\begin{tabular}{|c|c|c|c|}
\hline
\textbf{Modulation} & \textbf{$\tau = 0.7$} & \textbf{$\tau = 0.8$} & \textbf{$\tau = 0.9$} \\ \hline
BPSK                & --                  & 5\%                  & 46\%                   \\ \hline
QPSK                & 60\%                  & 74\%                  & 84\%                  \\ \hline
\end{tabular}
\end{table}

In addition to the earlier discussion comparing the computational complexity of the proposed CNN-based FTN detector with the M-BCJR algorithm, we also evaluated its computational efficiency against the methods presented in \cite{sugiura2013}, \cite{bedeer2017}, \cite{song2020}, and \cite{defilippo2025}. Notably, the algorithms in \cite{sugiura2013}, \cite{song2020}, and \cite{defilippo2025} require 8194 additions and 8196 multiplications, 9720 additions and 9720 multiplications, and 53,500 MAC operations, respectively, for the detection of a single symbol. These computational requirements clearly indicate that the complexity of these methods exceeds that of the proposed CNN-FK detector. Therefore, we selected the method presented in \cite{bedeer2017} for a more detailed analysis of computational complexity, as it provides an appropriate reference for comparison with the proposed CNN-FK detector. This comparison was conducted for \(\tau = 0.8\) under BPSK modulation, with the operation numbers summarized in Table~\ref{tab:comp_op_goback}. Additionally, the overall LUT-weighted computational cost metric for the Go-Back-\textit{K} algorithm is calculated as 1,696,790. When compared to the corresponding metric for the proposed method, which is 1,040,900, the analysis reveals that the proposed CNN-based method is approximately 39\% less complex. This significant reduction highlights the resource efficiency of the proposed method, even when benchmarked against another low-complexity algorithm. Furthermore, as discussed in Section~\ref{subsec:ber_results}, the BER performance of the proposed method also surpasses that of \cite{bedeer2017} for \(\tau = 0.8\) and \(\tau = 0.7\), demonstrating its robustness in detecting FTN signaling while maintaining computational efficiency. Together, these results position the proposed method as a compelling alternative for scenarios requiring both high detection accuracy and reduced computational cost.

\begin{table}[h!]
\centering
\caption{Comparison of Operation Numbers for Go-Back-\textit{K} and CNN-Based FTN Detector (100 Symbols, BPSK, \(\tau=0.8\))}
\resizebox{\columnwidth}{!}{%
\begin{tabular}{@{}lccccc@{}}
\toprule
\textbf{Algorithm} & \textbf{Multiplication} & \textbf{Addition} & \textbf{Comparison} & \textbf{Tanh} & \textbf{Sigmoid} \\
\midrule
Go-Back-\textit{K} & 14850  & 15444  & 300  & --   & --  \\
CNN                & 8100   & 12400  & --   & 1500 & 100 \\
\bottomrule
\end{tabular}%
}
\label{tab:comp_op_goback}
\end{table}

\section{Conclusion}
\label{sec:conclusion}
In conclusion, this study introduces a novel CNN-based FTN detection approach leveraging structured fixed kernel layers with domain-informed masking. Unlike traditional CNNs with moving kernels, the proposed method employs fixed convolutional kernels to explicitly learn ISI patterns at varying distances. A hierarchical filter allocation strategy enhances feature representation while optimizing computational efficiency. The detector achieves near-optimal BER performance, comparable to the BCJR algorithm for $\tau \geq 0.7$, with computational gains of up to 46\% over M-BCJR for BPSK and 84\% for QPSK. It is also evaluated against various existing methods, demonstrating efficiency and reliability. Moreover, its applicability to higher-order modulations, including 64-QAM, confirms its scalability. To assess practical robustness, the proposed detector has also been evaluated under quasi-static multipath Rayleigh fading conditions, demonstrating reliable performance beyond AWGN scenarios and confirming its suitability for low-mobility wireless environments where ISI remains dominant.

Nevertheless, several promising directions remain open for future research. Expanding the evaluation to fiber-optic communication systems—where impairments such as chromatic dispersion and nonlinearities prevail—can provide further insights into the versatility of the proposed detector. Additionally, assessing performance in more dynamic wireless environments, including time-varying channels affected by Doppler shifts, Rician fading, or multi-user interference, would be valuable for practical deployment. Future work may also consider hardware-efficient implementations to meet real-time processing requirements. Lastly, exploring alternative network architectures, particularly RNNs, may enhance the detector’s capacity to model long-range dependencies and improve performance in severe ISI regimes, especially for $\tau < 0.7$.

\end{document}